\documentclass{article}

\usepackage[utf8]{inputenc} \usepackage[T1]{fontenc}    \usepackage{url}            \usepackage{booktabs}       \usepackage{amsfonts}       \usepackage{nicefrac}       \usepackage{microtype}      \usepackage{xcolor}         \usepackage{setspace}
\usepackage{geometry}
\usepackage{natbib}
 \usepackage[colorlinks=true,linkcolor=red,citecolor=blue]{hyperref}

\usepackage{etoolbox}
\ifundef{\abstract}{}{\patchcmd{\abstract}{\quotation}{\quotation\noindent\ignorespaces}{}{}}

\usepackage[parfill]{parskip}
\usepackage{dsfont}
\usepackage{threeparttable}
\usepackage{amssymb}
\usepackage{mathtools}
\usepackage{float}
\usepackage{upgreek}
\allowdisplaybreaks
\usepackage{bm}
\usepackage{textcomp}
\usepackage{enumitem}
\usepackage{multirow}
\usepackage{colortbl}
\usepackage{xcolor}
\usepackage{makecell}
\usepackage{microtype}
\usepackage{graphicx}
\usepackage{subfigure}
\usepackage{booktabs}
 \usepackage[linesnumbered,ruled,vlined]{algorithm2e}

\usepackage{stackengine}
\usepackage{amsthm}

\newcommand{\q}{q}

\newcommand{\p}{p}
\newcommand{\D}{\mathbb{D}}

\newcommand{\e}{\mathbb{E}}
\newcommand{\V}{\mathbb{V}}

\newcommand{\ind}{\perp \!\!\! \perp }

\allowdisplaybreaks

\newtheorem{lemma}{Lemma}
\newtheorem{assumption}{Assumption}

\usepackage{babel}
\usepackage[font=small,labelfont=bf]{caption}

\title{\textbf{Federated Estimation of Causal Effects\\from Observational Data}}

\author{\normalsize
  \bfseries{
  Thanh Vinh Vo\textsuperscript{\normalfont{ 1}}\qquad Trong Nghia Hoang\textsuperscript{\normalfont{ 2}}\thanks{This work has been done when Nghia Hoang was with the MIT-IBM Watson AI Lab.} \qquad Young Lee\textsuperscript{\normalfont{ 3}} 
  \qquad Tze-Yun Leong\textsuperscript{\normalfont{ 1}}}\\[0.3cm]
  \normalsize
  \textsuperscript{\normalfont{1}}National University of Singapore  \qquad \textsuperscript{\normalfont{2}}AWS AI Labs  \qquad \textsuperscript{\normalfont{3}}Harvard University\\
}

\date{ }

\begin{document}

\maketitle

\begin{abstract}
\setlength{\parindent}{0em}

Many modern applications collect data that comes in federated spirit, with data kept locally and undisclosed. Till date, most insight into the causal inference requires data to be stored in a central repository. We present a novel framework for causal inference with federated data sources. We assess and integrate local causal effects from different private data sources without centralizing them. Then, the treatment effects on subjects from observational data using a non-parametric reformulation of the classical potential outcomes framework is estimated. We model the potential outcomes as a random function distributed by Gaussian processes, whose defining parameters can be efficiently learned from multiple data sources, respecting privacy constraints. We demonstrate the promise and efficiency of the proposed approach through a set of simulated and real-world benchmark examples.

 \end{abstract}

\section{Introduction} 
\label{sec:intro}
Estimating the casual effects of an intervention on an outcome is commonly used in many practical areas, e.g., personalized medicine \citep{powers2018some}, digital experiments \citep{taddy2016nonparametric} and political science \citep{green2012modeling}. One example is to estimate the effect of smoking on causing lung cancer. To accurately infer the causal effects, one would need a large number of data observations. However, observational data often exist across different institutions and typically cannot be centralized for processing due to privacy constraints. For example, medical records of patients are kept strictly confidential at local hospitals \citep{gostin2009beyond}. This real-life scenario would limit access of causal inference algorithms on the training data.

Existing medical data have not been fully exploited by causal inference primarily because of the aforementioned constraints. Current approaches in causal inference \citep[e.g.,][]{shalit2017estimating,yao2018representation} require the medical records to be shared and put in one place for processing. This could violate the privacy rights of patients. Some alternative solutions such as establishing data use agreements or creating secure data environments may not be possible and is often not implemented. For example, suppose that some hospitals own the electronic health records (EHRs) of different patient populations and we wish to utilize these EHRs to perform causal inference on whether smoking causes lung cancer in all of these populations. However, these records cannot be shared across the hospitals because they may contain sensitive information of the patients. This problem would lead to a big barrier for developing effective causal effect estimators that are generalizable, which usually need a diverse and big dataset. How to utilize these EHRs to build a global causal effect estimator while preserving the patients' privacy rights is a challenging problem which has not been well explored.

In practice, it is intractable to verify whether the causal estimands are reliable. Thus, in addition to giving point estimates of causal effects, an estimator which outputs confidence intervals would give helpful insights into the uncertainty of causal estimands. For example, a narrow confidence interval for the individual treatment effect means that patients are at a higher risk of getting lung cancer. Most of the recent causal effect estimators \citep[e.g.][]{shalit2017estimating,louizos2017causal,yao2018representation,madras2019fairness}, however, ignore discussion on the uncertainty of the causal effects. Some existing causal inference packages such as \texttt{econml} \citep*{econml} provides frequentist approaches, e.g., Bootstrap \citep{efron1994introduction}, Bootstrap-of-Little-Bags \citep{kleiner2014scalable}, to find such confidence intervals. These approaches require many rounds of resampling the entire dataset and retraining the models on these resamples. Hence, to use these approaches for the context of muti-source causal inference while preserving privacy, it might require a careful redesigning of the resampling algorithms.

These challenges motivate us to propose a framework that can learn the causal effects of interest without combining data sources to a central site and, at the same time, learn higher-order statistics of the causal effects, hence capturing their uncertainty. To address such problem, we utilize the Bayesian imputation approach \citep[][]{imbens2015causal} since it can capture uncertainty of the causal estimands. We then generalize this model to a more generic model based on Gaussian processes (GPs). To train the model on multiple sources while preserving data privacy, we further decompose this generic model into multiple components, each of them handling a source in our multi-source data context. This generic approach is called federated learning which was introduced recently in \citet{mcmahan2017communication} and it has not been studied for causal inference.  In short, our contributions are as follows:

\begin{itemize}[topsep=0pt,leftmargin=*]
\item We propose a novel Federated Causal Inference (FedCI) framework that fuses federated learning and causal inference to incorporate multiple data sources while preserving private rights of users.
\item An advantage of the proposed method is that it also gives higher-order statistics of the causal estimands under a Bayesian approach. \item We propose a variational approximation scheme for the proposed model whose evidence lower bound can be decomposed additively across different data sources. This allows the parameters to be optimized via federated gradient averaging. We then leverage the computed predictive distribution to estimate the desired treatment effect quantities efficiently. We carry out an empirical evaluation of the proposed framework on benchmark datasets, which shows competitive performance compared to the baselines trained on the combined dataset.
\end{itemize}

\section{Background and related work}
\label{sec:potential-outcomes}

\textbf{Causal inference.} In  most causal inference literature,  the estimation of causal effects is performed directly on accessible local data sources. \citet{hill2011bayesian,Alaa:2017,Alaa:2018} proposed a nonparametric approaches to estimate causal effects. 
A growing literature, including \cite{shalit2017estimating,Yoon:2018ganite,yao2018representation,kunzel2019metalearners,nie2020quasi}, used parametric methods to model the potential outcomes. These methods make a standard ignorability assumption of \citet{rosenbaum1983central}.  \citet{louizos2017causal,madras2019fairness} followed the structural causal model (SCM) \citep{pearl1995causal} to estimate causal effects under the existence of latent confounding variables.  \citet{bica2020Estimating,bica2020time} formalized potential outcomes for temporal data with observed and unobserved confounding variables to estimate counterfactual outcomes for treatment plans. All these works were not proposed for the context of multi-source data which cannot be shared and combined as a unified dataset due to some privacy constraints. 
Our model, in contrast, learns individual treatment effect (ITE) and average treatment effect (ATE) while preserving privacy of the observed individuals. It is different from the problem of transportability of causal relations \citep[e.g.,][]{pearl2011transportability,bareinboim2013meta,bareinboim2013causal,bareinboim2016causal}, where theoretical tools were developed to transport causal effects from a source population to a target population and did not take into account the privacy constraints.

\textbf{Federated learning.} The concepts of federated learning and causal inference are two well-known areas that have been developed independently. 
Federated learning aims to train an algorithm across multiple decentralized clients, thus preserving the privacy information of the data \citep{mcmahan2017communication}. Two variations of federated learning include federated stochastic gradient descent \citep{shokri2015privacy} and federated averaging \citep{mcmahan2017communication}. Recent developments of these two areas, e.g., \citet{alvarez2019non,zhe2019scalable,de2020mogptk,joukov2020fast} and \citet{hard2018federated,zhao2018federated,sattler2019robust,mohri2019agnostic} are formalized for a typical classification or regression problem. Federated learning has recently been applied in facilitating multi-institutional collaborations without sharing patient data \citep{rieke2020future,sheller2020federated} and healthcare informatics \citep{lee2020federated,xu2021federated}. Several applications of federated learning in medical data include predicting hospitalizations for cardiac events \citep{brisimi2018federated}, predicting adverse drug reactions \citep{choudhury2019predicting}, stroke prevention \citep{ju2020privacy}, mortality prediction \citep{vaid52federated}, medical imaging \citep{ng2021federated}, predicting outcomes in SARS-COV-2 patients \citep{flores2021federated}. However, to the best of our knowledge, no work has been done for causal inference.

Following some recent works in causal inference \citep[e.g.,][]{shalit2017estimating,yao2018representation,oprescu2019orthogonal,kunzel2019metalearners,nie2020quasi}, we utilize the potential outcomes framework to develop a federated causal inference algorithm. 
Our approach has connection to the SCM approach with a causal graph that includes three variables: treatment, outcome, and observed confounder \citep[][Chapter~7]{pearl2009causality}, where the causal effects can be identified using backdoor adjustment formula \citep{pearl2009causality}. We summarize the related models in the subsequent sections.

\subsection{Potential outcomes}
The concept of potential outcomes was proposed in \citet{neyman1923application} for randomized trial experiments. \citet{Rubin:1975,rubin:1976a,rubin1977assignment,Rubin:1978} re-formalized the framework for observational studies. We consider the causal effects of a binary treatment $w$, with $w=1$ indicating assignment to `treatment' and $w=0$ indicating assignment to `control'. Following convention in the literature \citep[e.g.,][]{Rubin:1978}, the causal effect for individual $i$ is defined as a comparison of the two potential outcomes, $y_i(0)$ and $y_i(1)$, where these are the outcomes that would be observed under $w=1$ and $w=0$, respectively. We can never observe both $y_i(1)$ and $y_i(0)$ for any individual $i$, because it is not possible to go back in time and expose the $i$--th  individual to the other treatment. Therefore, individual causal effects cannot be known and must be inferred.

\subsection{Missing outcomes imputation}
\label{sec:rubin-imputation}
In this work, we generalize the Bayesian imputation model of \citet[][Chapter 8]{imbens2015causal} since this model can capture uncertainty of the causal estimands in a Bayesian setting. The model is specified as follows:
\begin{align}
    y_i(0) &= \bm{\upbeta}_0^\top\mathbf{x}_i + \epsilon_{0i}, &y_i(1) &= \bm{\upbeta}_1^\top\mathbf{x}_i + \epsilon_{1i},\label{eq:rubin-model}
\end{align}
where $\epsilon_{0i}$ and $\epsilon_{1i}$  are the Gaussian noises. The key to compute treatment effects is $y_i(0)$ and $y_i(1)$, however we cannot observe both of them. So we need to impute one of the two outcomes. Let $y_{i,\textrm{obs}}$ be the observed outcome and $y_{i,\textrm{mis}}$ be the unobserved outcome. The idea is to find the marginal distribution $\p(y_{i,\textrm{mis}}|\mathbf{y}_{\textrm{obs}},\mathbf{X},\mathbf{w})$. Once the missing outcomes are imputed, the treatment effects can be estimated. Note that $\p(y_{i,\textrm{mis}}|\mathbf{y}_{\textrm{obs}},\mathbf{X},\mathbf{w}) \neq \p(y_{i,\textrm{mis}}|y_{i,\textrm{obs}},\mathbf{x}_i,w_i)$, i.e., the outcomes of all individual are dependent. 
To find the above above distribution, \citet[][]{imbens2015causal} suggested four steps based on the following equation $
    \p(y_{i,\textrm{mis}}| \mathbf{y}_{\textrm{obs}},\mathbf{X},\mathbf{w}) = \int \p(y_{i,\textrm{mis}}| \mathbf{y}_{\textrm{obs}},\mathbf{X},\mathbf{w},\theta)\p(\theta|\mathbf{y}_{\textrm{obs}},\mathbf{X},\mathbf{w})d\theta$ where $\theta$ is the set of all parameters in the model, i.e., $\theta = \{\bm{\upbeta}_0, \bm{\upbeta}_1\}$. The aim is to find $\p(y_{i,\textrm{mis}}| \mathbf{y}_{\textrm{obs}},\mathbf{X},\mathbf{w},\theta)$ and $\p(\theta|\mathbf{y}_{\textrm{obs}},\mathbf{X},\mathbf{w})$, and then compute the above integration to obtain  $\p(y_{i,\textrm{mis}}| \mathbf{y}_{\textrm{obs}},\mathbf{X},\mathbf{w})$, which is a non-parametric prediction. In Sections~\ref{sec:rubin-to-gp}, \ref{sec:model} and \ref{sec:inference}, we generalize this model with Gaussian processes and decompose it into multiple components to perform federated inference of the causal effects. 

\section{Federated causal model}

This section formalizes the problem of estimating causal effects under some privacy constraints. 
We address this problem by generalizing the Bayesian imputation model presented in Section~\ref{sec:rubin-imputation} to a more generic model based on Gaussian processes. We decompose the model into multiple components, each associated with a data source. This decomposition results in the proposed Federated Causal Inference (FedCI) method.

\subsection{Problem formulation}
\label{sec:prob-formu}
In the following, we detail our proposed model specification and explicate the link to the causal quantity that we would like to estimate. 

\textbf{Problem setting \& notations.} Suppose we have $m$ sources of data, each is denoted by $\mathsf{D}^\mathsf{s} = \{( w_i^\mathsf{s}, y_{i,\textrm{obs}}^\mathsf{s}, \mathbf{x}_i^\mathsf{s})\}_{i=1}^{n_\mathsf{s}}$, where $\mathsf{s}=1,2,\dots,m$, and the quantities $w_i^\mathsf{s}$, $y_{i,\textrm{obs}}^\mathsf{s}$ and  $\mathbf{x}_i^\mathsf{s}$ are the treatment assignment, observed outcome associated with the treatment, and covariates of individual $i$ in source $\mathsf{s}$, respectively. In this work, we focus on binary treatment $w_i^\mathsf{s} \in \{0,1\}$, thus $y_{i,\textrm{obs}}^\mathsf{s}$ can be either the potential outcomes $y_i^\mathsf{s}(0)$ or $y_i^\mathsf{s}(1)$, i.e., for each individual $i$, we can only observe either $y_i^\mathsf{s}(0)$ or  $y_i^\mathsf{s}(1)$, but not both of them. We further denote the unobserved or missing outcome as $y_{i,\textrm{mis}}^\mathsf{s}$. These variables are related to each other through the following equations
\begin{align}
y_i^\mathsf{s}(1) &= w_i^\mathsf{s} y^\mathsf{s}_{i,\textrm{obs}} + (1-w_i^\mathsf{s})y^\mathsf{s}_{i,\textrm{mis}}, &y_i^\mathsf{s}(0) &= (1-w_i^\mathsf{s}) y^\mathsf{s}_{i,\textrm{obs}} + w_i^\mathsf{s}y^\mathsf{s}_{i,\textrm{mis}}\label{eq:y0-y1-ymis-yobs}.
\end{align}
Thus, $y_i^\mathsf{s}(1) = y^\mathsf{s}_{i,\textrm{obs}}$ when $w_i^\mathsf{s}=1$ and $y_i^\mathsf{s}(1) = y^\mathsf{s}_{i,\textrm{mis}}$ when $w_i^\mathsf{s}=0$, and similar for $y_i^\mathsf{s}(0)$. 
For notational convenience, we further denote
\begin{align*}
    \mathbf{y}^\mathsf{s}(0) &= [y_1^\mathsf{s}(0),\!...,y_{n_\mathsf{s}}^\mathsf{s}(0)]^\top,  &\mathbf{y}^\mathsf{s}_{\textrm{obs}} &= [y^\mathsf{s}_{1,\textrm{obs}},\!...,y^\mathsf{s}_{n_\mathsf{s},\textrm{obs}}]^\top,
\end{align*}
and similarly for $\mathbf{y}^\mathsf{s}(1)$, $\mathbf{y}^\mathsf{s}_{\textrm{mis}}$, $\mathbf{X}^\mathsf{s}$ and $\mathbf{w}^\mathsf{s}$.

\textbf{Causal effects of interest.} Due to privacy concerns, these data sources $\mathsf{D}^\mathsf{s}$ are located in different locations. We are interested in estimating individual treatment effect (\textrm{ITE}) and average treatment effect (\textrm{ATE}) which are defined as follows
\begin{align}
\uptau_i^\mathsf{s} &\vcentcolon= y_i^\mathsf{s}(1) - y_i^\mathsf{s}(0), &\uptau &\vcentcolon= \Big(\sum_{\mathsf{s}=1}^m\sum_{i=1}^{n_\mathsf{s}}\uptau_i^\mathsf{s}\Big)/n,\label{eq:ate}
\end{align}
where $y_i^\mathsf{s}(1)$ and $y_i^\mathsf{s}(0)$ are realization outcomes of their corresponding random variables, and $n = \sum_{\mathsf{s}=1}^m n_\mathsf{s}$ is the total number of samples. Note that the ITE is also known as the conditional average treatment effect (CATE). 
\subsection{The causal estimands}
Inserting Eq.~(\ref{eq:y0-y1-ymis-yobs}) into (\ref{eq:ate}), we obtain the estimate of ITE
\begin{align}
&\e[\uptau^\mathsf{s}_i]
=\tilde{w}_i^\mathsf{s} (y^\mathsf{s}_{i,\textrm{obs}} - \e\big[y^\mathsf{s}_{i,\textrm{mis}}\big| \mathbf{y}_{\textrm{obs}}, \mathbf{X}, \mathbf{w}\big]), 
&\V\text{ar}[\uptau^\mathsf{s}_i]&=(\tilde{w}_i^\mathsf{s})^2\mathbb{V} \text{ar}\left[y^\mathsf{s}_{i,\textrm{mis}}\big| \mathbf{y}_{\textrm{obs}}, \mathbf{X}, \mathbf{w}\right],\label{eq:ite-hat-expectation}
\end{align}
where $\tilde{w}_i^\mathsf{s} \vcentcolon= 2w_i^\mathsf{s}-1$ and $\mathbf{y}_{\textrm{obs}}$, $\mathbf{X}$, $\mathbf{w}$ denotes the vectors/matrices of the observed outcomes, covariates and treatments concatenated from all the sources. The estimate of ATE is as follows
\begin{align}
&\e[\uptau] =\mathbf{\tilde{w}}^\top(\mathbf{y}_{\textrm{obs}} - \e[\mathbf{y}_{\text{mis}}\,|\, \mathbf{y}_{\textrm{obs}}, \mathbf{X}, \mathbf{w}])/n, &\V\text{ar}[\uptau]&=\mathbf{\tilde{w}}^\top\mathbb{C} \text{ov}[\mathbf{y}_{\textrm{mis}}\,|\, \mathbf{y}_{\textrm{obs}}, \mathbf{X}, \mathbf{w}]\mathbf{\tilde{w}}/n^2,\label{eq:tau-hat-variance}
\end{align}
where $\mathbf{\tilde{w}} \vcentcolon=2\mathbf{w}-\mathbf{1}$ with $\mathbf{1}$ is a vector of ones. The above estimates capture the mean and variance of the treatment effects. At present, what remains is to learn the posterior $\p(\mathbf{y}_{\textrm{mis}}\big| \mathbf{y}_{\textrm{obs}}, \mathbf{X}, \mathbf{w})$, which is the predictive distribution of $\mathbf{y}_{\textrm{mis}}$ given \emph{all} the covariates, treatments and observed outcomes from \emph{all} sources. In the next sections, we develop a federated GP-augmented imputation model to approximate this distribution.

\subsection{Assumptions}
\label{sec:assumptions}
\begin{figure}\centering
    \includegraphics[width=0.8\textwidth]{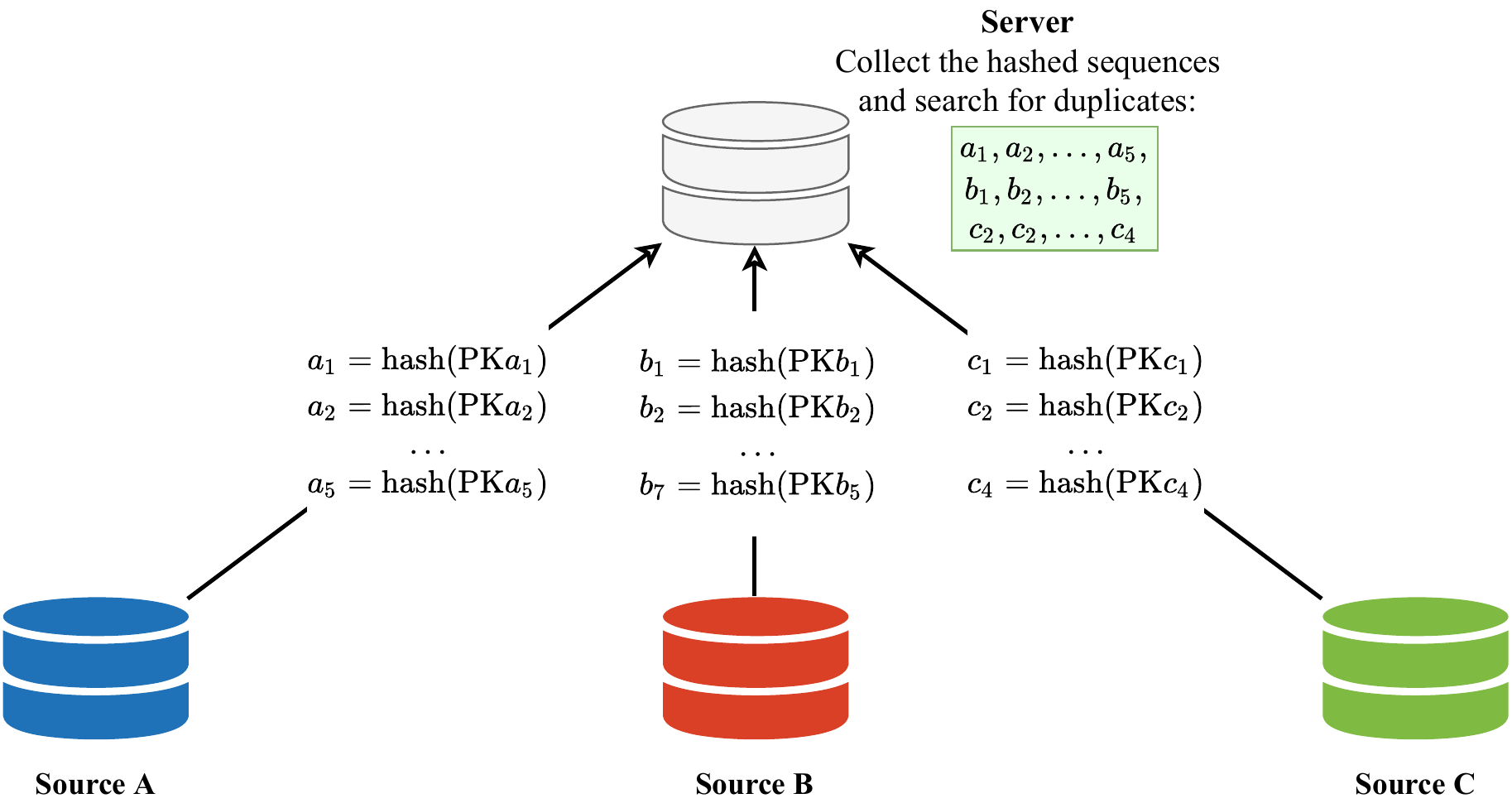}
    \caption{The secure preprocessing procedure to identify duplicated individuals among multiple sources. $\text{PK}a_i$ ($i=1,\!...,5$), $\text{PK}b_i$ ($i=1,\!...,7$), $\text{PK}c_i$ ($i=1,\!...,4$) are the primary keys of each individual in each source. $a_i$ ($i=1,\!...,5$), $b_i$ ($i=1,\!...,7$), $c_i$ ($i=1,\!...,4$) are the hashed sequences of these individuals.}
    \label{fig:preprocessing}
    \vspace{-6pt}
\end{figure}

In the following, we make some assumptions that allow the causal effects to be estimate in a federated setting. The first three assumptions are standard. The fourth assumption is needed to allow us to proceed in the preprocessing step.
\begin{assumption}
[Unconfoundedness] \label{assumption:ignorability} $ y_i^\mathsf{s}(1),y_i^\mathsf{s}(0)\ind w_i^\mathsf{s}\,|\,\mathbf{x}_i^\mathsf{s}$. \citep{rosenbaum1983central}
\end{assumption}
\begin{assumption} [The stable unit treatment value assumption] \label{assumption:sutva} \emph{($i$)} There are no hidden versions of the treatment and \emph{($ii$)} treatment on one unit does not affect the potential outcomes of another one. \citep{imbens2015causal}
\end{assumption}
\begin{assumption} \label{assumption:share-covariates}
The individuals from all sources share the same set of covariates.
\end{assumption}
\begin{assumption}\label{assumption:unique-ident} There exists a set of features such that any individual is uniquely identified across different sources. We refer to this set as `primary key'.
\end{assumption}
A `primary key' in Assumption~\ref{assumption:unique-ident} is not limited to the observed data used for inference as described in Section~\ref{sec:prob-formu}, but it can be any features to uniquely identify an individual such as $\{\text{nationality, national id}\}$ of patients. 
Assumption~\ref{assumption:unique-ident} allows us to proceed with a preprocessing procedure (if necessary) to remove the repeated individuals in different sources while preserving the individuals' privacy. The preprocessing procedure are summarized as follows. Firstly, each source would use a one-way hash function (such as MD4, MD5, SHA or SHA256) to encrypt each individuals' primary key and then send the hashed sequences to a server. By doing this, the individuals' data are secured. Note that the one-way hash function is agreed among the sources so that they would use the same function. Then, the server collects all hashed sequences from all sources and perform a matching algorithm to see if there exists repeated individuals among different sources. For each repeated individual, the server randomly choose to keep it on a small number (predefined) of sources and inform the other sources to exclude this individual from the training process. The whole procedure is to ensure that an individual does not exists in a huge number of sources, thus prevent learning a biased model. The whole procedure is to ensure that an individual does not exists in a huge number of sources, thus prevent learning a biased model. We summarize the procedure in Figure~\ref{fig:preprocessing}.

Assumption~\ref{assumption:unique-ident} and the preprocessing procedure are required for data that are highly repeated in different sources only. For data that are not likely to have a high number of repetitions such as patients from different hospitals of different countries, the above assumption and the preprocessing procedure are not required. Note that the existing methods also need Assumption~\ref{assumption:unique-ident} since they need to combine data and remove repeated individuals.

Note that Assumption~\ref{assumption:ignorability} is not testable since we cannot observe both $y_i^\mathsf{s}(0)$ and $y_i^\mathsf{s}(1)$, and this is well documented \citep{imai2010general}. However, Assumption~\ref{assumption:sutva} is likely to hold in a real-life setting. For example, a patient having an increase of blood pressure due under a medication cannot in any shape or form influence the blood pressure (outcome) of another patient. In addition, most hospitals should collect common covariates of their patients, thus Assumption~\ref{assumption:share-covariates} is also a reasonable assumption. By preceeding discussions, Assumption~\ref{assumption:unique-ident} is a realistic assumption. In the subsequent sections, we assume that all of the assumptions described in this section are satisfied, and the preprocessing procedure was performed if it is necessary.

\subsection{GP-based imputation}
\label{sec:rubin-to-gp}
The model presented in Eq.~(\ref{eq:rubin-model}) is a simple Bayesian linear model. In this section, we present a more generic nonlinear model under the Bayesian setting. In particular, since $\bm{\upbeta}_0$ and $\bm{\upbeta}_1$ follows multivariate normal distributions, the two components $\bm{\upbeta}_0^\top\mathbf{x}_i$ and $\bm{\upbeta}_1^\top\mathbf{x}_i$ also follow multivariate normal distributions. The generalisation of these two components are $f_0(\mathbf{x}_i) = \bm{\upbeta}_0^\top\omega(\mathbf{x}_i)$ and $f_1(\mathbf{x}_i) = \bm{\upbeta}_1^\top\omega(\mathbf{x}_i)$, where $\omega(\mathbf{x}_i)$ is a vector of basis functions with input $\mathbf{x}_i$. This formulation would lead to the fact that the marginal of $f_0(\mathbf{x})$ and $f_1(\mathbf{x})$ are Gaussian processes. Thus, we propose\begin{align}
    \!\!\!\!\!y_i(0) &= f_0(\mathbf{x}_i) + \epsilon_{0i}, &y_i(1) &= f_1(\mathbf{x}_i) + \epsilon_{1i},\label{eq:gp-model}
\end{align}
where $f_0(\mathbf{x}_i)$ and $f_1(\mathbf{x}_i)$ are two random functions evaluated at $\mathbf{x}_i$, i.e., $f_0(\mathbf{x}_i) \sim \mathsf{GP}(\mu_0(\mathbf{X}), \mathbf{K})$ and $f_1(\mathbf{x}_i) \sim \mathsf{GP}(\mu_1(\mathbf{X}), \mathbf{K})$, where $\mathbf{K}$ denotes the covariance matrix computed with a kernel function $\mathsf{k}(\mathbf{x},\mathbf{x}')$. Similar to the imputation model of \citet{imbens2015causal}, the model presented here also requires finding the marginal distribution $\p(y_{i,\textrm{mis}}\,|\,\mathbf{y}_{\textrm{obs}},\mathbf{X},\mathbf{w})$. Although this model is generic, it requires access to all of the observed data to compute $\mathbf{K}$, hence violates privacy rights. In the subsequent sections, we propose a federated model that address this problem. 

\subsection{The proposed model}
\label{sec:model}
Recall that the aim is to find $\p(\mathbf{y}_{\textrm{mis}}\,|\, \mathbf{y}_{\textrm{obs}}, \mathbf{X}, \mathbf{w})$ so that we may in turn compute Eqs.~(\ref{eq:ite-hat-expectation})~and~(\ref{eq:tau-hat-variance}) to arrive at the quantities of interest. To that end, we propose to model the joint distribution of the potential outcomes as follows
\begin{align}
\Aboxed{\begin{bmatrix}
	y_i^\mathsf{s}(0)\\
	y_i^\mathsf{s}(1)
	\end{bmatrix} = \Phi^{\frac{1}{2}}\left(\begin{bmatrix}
	f_0^\mathsf{s}(\mathbf{x}_i)\\
	f_1^\mathsf{s}(\mathbf{x}_i)
	\end{bmatrix} + \begin{bmatrix}
	g_0^\mathsf{s}\\
	g_1^\mathsf{s}
	\end{bmatrix}\right) + \Sigma^{\frac{1}{2}}\bm{\upvarepsilon}_i^\mathsf{s},}
	\label{eq:the-model}
\end{align}
where $\bm{\upvarepsilon}_i^\mathsf{s} \sim \mathsf{N}(\mathbf{0}, \mathbf{I}_2)$  is to handle the noise of the outcomes. 
As mentioned earlier in Section~\ref{sec:rubin-imputation}~and~\ref{sec:rubin-to-gp}, all the outcomes are \emph{dependent} in the Bayesian imputation approach. This dependency is handle via $f_j^\mathsf{s}(\mathbf{x}_i)$ and $g_j^\mathsf{s}$ ($j\in\{0,1\}$). We name the dependency handled by $f_j^\mathsf{s}(\mathbf{x}_i)$ as intra-dependency and the one captured by $g_j^\mathsf{s}$ as inter-dependency. 

\textbf{{\scriptsize$\blacksquare$} Intra-dependency.} 
$f_0^\mathsf{s}(\mathbf{x}_i)$ and $f_1^\mathsf{s}(\mathbf{x}_i)$ are GP-distributed functions, which allows us to model each source dataset simultaneously along with their heterogeneous correlation. Specifically, we model $f_0^\mathsf{s}(\mathbf{x}_i) \sim \mathsf{GP}(\mu_0(\mathbf{X}^\mathsf{s}), \mathbf{K}^\mathsf{s})$ and $f_1^\mathsf{s}(\mathbf{x}_i) \sim \mathsf{GP}(\mu_1(\mathbf{X}^\mathsf{s}), \mathbf{K}^\mathsf{s})$, where $\mathbf{K}^\mathsf{s}$ is a covariance matrix computed by a kernel function $\mathsf{k}(\mathbf{x}_i^\mathsf{s}, \mathbf{x}_j^\mathsf{s})$, and $\mu_0(\cdot)$, $\mu_1(\cdot)$ are functions modelling the mean of these GPs. Parameters of these functions and hyperparameters in the kernel function are shared across multiple sources. 
The above GPs handle the correlation within one source only. 

\textbf{{\scriptsize$\blacksquare$} Inter-dependency.} 
To capture \textit{dependency} among the sources, we introduce variable $\mathbf{g} = [\mathbf{g}_0, \mathbf{g}_1]$, where 
\begin{align*}
    \mathbf{g}_0 &= [g_0^{1},\!...,g_0^{m}]^\top \sim \mathsf{N}(\bm{r}_0, \mathbf{M}), &\mathbf{g}_1 &= [g_1^{1},\!...,g_1^{m}]^\top \sim \mathsf{N}(\bm{r}_1, \mathbf{M}).\end{align*}
Each $g_0^\mathsf{s}$ and $g_1^\mathsf{s}$ are shared within the source $\mathsf{s}$, and they are correlated across multiple sources $\mathsf{s} \in \{1,\!...,m\}$. The correlation among the sources is modelled via the covariance matrix $\mathbf{M}$ 
which is computed with a kernel function.  The inputs to the kernel function are the sufficient statistics (we used mean, variance, skewness, and kurtosis) of each covariate $\mathbf{x}^\mathsf{s}$ within the source $\mathsf{s}$. We denote the first four moments of covariates as $\mathbf{\tilde{x}}^\mathsf{s} \in \mathbb{R}^{4 d_x \times 1}$ and the kernel function as $\gamma(\mathbf{\tilde{x}}^\mathsf{s}, \mathbf{\tilde{x}}^{\mathsf{s}'})$, which evaluates the correlation of two source $\mathsf{s}$ and $\mathsf{s}'$. The above formulation implies that $\mathbf{g}_0$ and $\mathbf{g}_1$ are GPs. Each element of $\bm{r}_0$ and $\bm{r}_1$ are computed with the mean functions $r_0(\mathbf{\tilde{x}}^\mathsf{s})$ and $r_1(\mathbf{\tilde{x}}^\mathsf{s})$, respectively. In this setting, we only share the sufficient statistics of covariates, but not covariates of a specific individual, hence preserving privacy of all individuals.

\textbf{{\scriptsize$\blacksquare$} The two variables $\Phi$ and $\Sigma$.} 
These variables are positive semi-definite matrices capturing the correlation between the two possible outcomes $y_i^\mathsf{s}(0)$ and $y_i^\mathsf{s}(1)$, $\Phi^{\frac{1}{2}}$ and $\Sigma^{\frac{1}{2}}$ are their Cholesky decomposition matrices. Note that $\Phi$ and $\Sigma$ are also random variables. The reason that we constraint $\Phi$ and $\Sigma$ as positive semi-definite matrices is explained later in Lemma~\ref{lem:joint-prob-s}. Because of this constraint, we model their priors using Wishart distribution $\Phi  \sim \mathsf{Wishart}(\mathbf{V}_0, d_0)$, $\Sigma \sim \mathsf{Wishart}(\mathbf{S}_0, n_0)$, where $\mathbf{V}_0, \mathbf{S}_0 \in \mathbb{R}^{2\times 2}$ are predefined positive semi-definite matrices and $d_0, n_0 \ge 2$ are predefined degrees of freedom.

\begin{figure}\centering
		\includegraphics[width=0.45\textwidth]{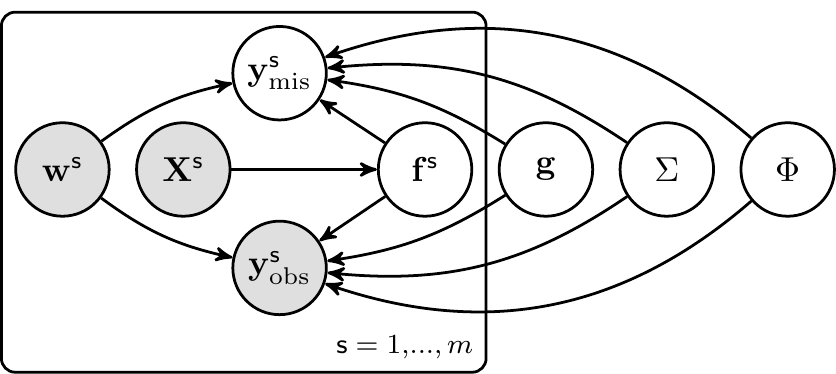}

    \caption{
        Graphical model that summarizes the proposed framework with treatment $\mathbf{w}^\mathsf{s}$, covariate $\mathbf{X}^\mathsf{s}$, and the two potential outcomes $\mathbf{y}_\textrm{mis}^\mathsf{s}$ and $\mathbf{y}_\textrm{obs}^\mathsf{s}$. The quantity $\mathbf{f}^\mathsf{s}$ is idiosyncratic to the sources and $\mathbf{g}$ contains shared characteristics across all the sources. $\Sigma$ and $\Phi$ are shared parameters. Note that this is not a causal graph.
} \label{fig:the-model}
  \vskip -9pt
\end{figure}

\textbf{{\scriptsize$\blacksquare$} The graphical model of our framework.} 
We summarize our framework in Figure~\ref{fig:the-model}. The figure shows that $\mathbf{g}$, $\Sigma$ and $\Phi$ are shared crosses the sources, thus capturing the correlation among them, and $\mathbf{f}^\mathsf{s}$ is specific for the source $\mathsf{s}$ that capture the correlation among individuals within this source.
To see how our model handles dependency between the outcomes of two different sources through the latent variable $\mathbf{g}$, we block the paths between two sources $\mathsf{s}$ and $\mathsf{s}'$ through $\Phi$ and $\Sigma$ and only keep the path through $\mathbf{g}$. The covariance between the outcomes of $\mathsf{s}$ and $\mathsf{s}'$ is presented in Lemma~\ref{lem:cov-cross-source}. 
\begin{lemma}
\label{lem:cov-cross-source}
Let $\mathsf{s}$ and $\mathsf{s}'$ be two different sources. Then, $\mathbb{C}\emph{\text{ov}}(\mathbf{y}_i^\mathsf{s}, \mathbf{y}_j^{\mathsf{s}'}\,|\,\Sigma,\Phi) = \Phi^{\frac{1}{2}}\Lambda^{(\mathsf{s},\mathsf{s}')}(\Phi^{\frac{1}{2}})^\top$, where $\Lambda^{(\mathsf{s},\mathsf{s}')} = \mathsf{diag}([\gamma(\mathbf{\tilde{x}}^\mathsf{s}, \mathbf{\tilde{x}}^{\mathsf{s}'}), \gamma(\mathbf{\tilde{x}}^\mathsf{s}, \mathbf{\tilde{x}}^{\mathsf{s}'})])$, $\mathbf{y}_i^\mathsf{s} = [y_i^\mathsf{s}(0), y_i^\mathsf{s}(1)]^\top$, and $\mathbf{y}_j^{\mathsf{s}'} = [y_j^{\mathsf{s}'}(0), y_j^{\mathsf{s}'}(1)]^\top$.
\end{lemma}
The diagonal of $\Phi^{\frac{1}{2}}\Lambda^{(\mathsf{s},\mathsf{s}')}(\Phi^{\frac{1}{2}})^\top$ in Lemma~\ref{lem:cov-cross-source} is non-zeros, which implies that $y_i^\mathsf{s}(0)$ and $y_j^{\mathsf{s}'}(0)$ are correlated, and so do $y_i^\mathsf{s}(1)$ and $y_j^{\mathsf{s}'}(1)$. 

\subsection{The proposed algorithm}
\label{sec:inference}
In this section, we present some results on the joint distribution of potential outcomes. Then, we construct an objective function that can be trained in a federated fashion.
\subsubsection{The joint distribution of the outcomes}

In the following, we derive some results that are helpful in constructing the federated objective function in Section~\ref{sec:objective-function}. Due to limited space, we defer the proofs of these results to Appendix. For convenience in presenting the subsequent results, we further denote $\mathbf{g}^\mathsf{s} = [\mathbf{g}_0^\mathsf{s}, \mathbf{g}_1^\mathsf{s}]$, where $\mathbf{g}_0^\mathsf{s} = [g_0^\mathsf{s},\!..., g_0^\mathsf{s}]^\top$ and $\mathbf{g}_1^\mathsf{s} = [g_1^\mathsf{s},\!..., g_1^\mathsf{s}]^\top$. 

\begin{lemma}
\label{lem:joint-prob-s}
Let $\Phi$, $\Sigma$, $\mathbf{K}$, $\mu_0(\mathbf{X}^\mathsf{s})$, $\mu_1(\mathbf{X}^\mathsf{s})$, and $\mathbf{g}^\mathsf{s}$  satisfy the model in Eq.~\emph{(\ref{eq:the-model})}. Then,
\begin{align*}
&\begin{bmatrix}
\mathbf{y}^\mathsf{s}(0)\\
\mathbf{y}^\mathsf{s}(1)
\end{bmatrix}\Big|\Phi, \Sigma, \mathbf{X}^\mathsf{s}, \mathbf{w}^\mathsf{s}, \mathbf{g}^\mathsf{s} \sim \mathsf{N}\!\left( \left(\Phi^{\frac{1}{2}} \otimes \mathbf{I}_{n_\mathsf{s}}\right)\begin{bmatrix}\mu_0(\mathbf{X}^\mathsf{s}) + \mathbf{g}_0^\mathsf{s}\\\mu_1(\mathbf{X}^\mathsf{s}) + \mathbf{g}_1^\mathsf{s}
\end{bmatrix} \!\!, \Phi \otimes \mathbf{K}^\mathsf{s} + \Sigma \otimes \mathbf{I}_{n_\mathsf{s}}\right)\!\!,
\end{align*}
where $\otimes$ is the Kronecker product.
\end{lemma}
From Lemma~\ref{lem:joint-prob-s}, we observe that $\Phi$, $\mathbf{K}^\mathsf{s}$, $\Sigma$, and $\mathbf{I}_{n_\mathsf{s}}$ are positive semi-definite, thus the covariance matrix $\Phi \otimes \mathbf{K}^\mathsf{s} + \Sigma \otimes \mathbf{I}_{n_\mathsf{s}}$ is positive semi-definite due to the fundamental property of Kronecker product. This explains the reason we chose $\Phi$ and $\Sigma$ to be positive semi-definite in our model; otherwise, the covariance matrix is invalid. From Lemma~\ref{lem:joint-prob-s}, we can obtain the following result in Lemma~\ref{lem:3}.
\begin{lemma}
\label{lem:3}
Let $\Phi$, $\Sigma$, $\mathbf{K}$, $\mu_0(\mathbf{X}^\mathsf{s})$, $\mu_1(\mathbf{X}^\mathsf{s})$, and $\mathbf{g}^\mathsf{s}$  satisfy the model in Eq.~\emph{(\ref{eq:the-model})}. Then,
\begin{align*}
&\!\!\!\begin{bmatrix}
\mathbf{y}^\mathsf{s}_{\emph{\textrm{obs}}}\\
\mathbf{y}^\mathsf{s}_{\emph{\textrm{mis}}}
\end{bmatrix}\Big|\Phi, \Sigma, \mathbf{X}^\mathsf{s}, \mathbf{w}^\mathsf{s}, \mathbf{g}^\mathsf{s} \sim \mathsf{N}\!\left( \begin{bmatrix}\mu_{\emph{\textrm{obs}}}(\mathbf{X}^\mathsf{s})\\\mu_{\emph{\textrm{mis}}}(\mathbf{X}^\mathsf{s})
\end{bmatrix} \!\!,\! \begin{bmatrix}\mathbf{K}_{\emph{\textrm{obs}}}^\mathsf{s}&\mathbf{K}_{\emph{\textrm{om}}}^\mathsf{s}\\
(\mathbf{K}_{\emph{\textrm{om}}}^\mathsf{s})^\top&\mathbf{K}_{\emph{\textrm{mis}}}^\mathsf{s}\end{bmatrix}\right)\!\!,
\end{align*}
The mean functions $\mu_{\emph{\textrm{obs}}}(\mathbf{X}^\mathsf{s})$ and $\mu_{\emph{\textrm{mis}}}(\mathbf{X}^\mathsf{s})$ are as follows:
\begin{align*}
\mu_{\emph{\textrm{obs}}}(\mathbf{X}^\mathsf{s}) &= (\mathbf{1} - \mathbf{w}^\mathsf{s})\odot\mathbf{m}_0 + \mathbf{w}^\mathsf{s} \odot\mathbf{m}_1, &\mu_{\emph{\textrm{mis}}}(\mathbf{X}^\mathsf{s}) &= \mathbf{w}^\mathsf{s} \odot\mathbf{m}_0 + (\mathbf{1} - \mathbf{w}^\mathsf{s}) \odot\mathbf{m}_1,
\end{align*}
where $\mathbf{m}_0 = \phi_{11}^\ast(\mu_0(\mathbf{X}^\mathsf{s}) + \mathbf{g}_0^\mathsf{s})$ and $\mathbf{m}_1 = \phi_{21}^\ast(\mu_0(\mathbf{X}^\mathsf{s}) + \mathbf{g}_0^\mathsf{s}) + \phi_{22}^\ast(\mu_1(\mathbf{X}^\mathsf{s}) + \mathbf{g}_1^\mathsf{s})$ with $\phi^\ast_{ab}$ is the $(a,b)$--th element of Cholesky decomposition matrix of $\Phi$, $\mathbf{1}$ is a vector ones, and $\odot$ is the element-wise product. The covariance matrices $\mathbf{K}^\mathsf{s}_{\textrm{\emph{\textrm{obs}}}}$, $\mathbf{K}^\mathsf{s}_{\textrm{\emph{\textrm{mis}}}}$, and $\mathbf{K}^\mathsf{s}_{\textrm{\emph{\textrm{om}}}}$ are computed by kernel functions:
\begin{align*}
k_{\emph{\textrm{obs}}}(\mathbf{x}_i, \mathbf{x}_j) \!&=\! \big[(1-w_i)(1-w_j)\phi_{11} + w_iw_j\phi_{22} + (1-w_i)w_j\phi_{12} + w_i(1-w_j)\phi_{21}\big] \mathsf{k}(\mathbf{x}_i, \mathbf{x}_j)\\[-0.1cm]
&\,\,\,\,\, + \big[(1-w_i)\sigma_{11} + w_i\sigma_{22}\big] \mathds{1}_{i=j},\\[-0.1cm]
k_{\emph{\textrm{mis}}}(\mathbf{x}_i, \mathbf{x}_j) \!&=\! \big[w_iw_j\phi_{11} + (1-w_i)(1-w_j)\phi_{22} \,\,\!+\! (1-w_i)w_j\phi_{21} + w_i(1-w_j)\phi_{12}\big] \mathsf{k}(\mathbf{x}_i, \mathbf{x}_j)\\[-0.1cm]
& \,\,\,\,\,+ \big[w_i\sigma_{11} + (1-w_i)\sigma_{22}\big] \mathds{1}_{i=j},
\\[-0.1cm]
k_{\emph{\textrm{om}}}(\mathbf{x}_i, \mathbf{x}_j) &= \big[(1-w_i)(1-w_j)\phi_{21} + w_iw_j\phi_{12}+ (1-w_i)w_j\phi_{22} + w_i(1-w_j)\phi_{11}\big] \mathsf{k}(\mathbf{x}_i, \mathbf{x}_j) \\[-0.1cm]
&\,\,\,\,\,+ \big[(1-w_i)\sigma_{21} + w_i\sigma_{12}\big] \mathds{1}_{i=j},
\end{align*}
where $\phi_{ab}$ and $\sigma_{ab}$ are the $(a,b)$--th elements of $\Phi$ and $\Sigma$, respectively.
\end{lemma}

In the subsequent sections, we use the result in Lemma~\ref{lem:3} to obtain the conditional likelihood $\p(\mathbf{y}^\mathsf{s}_{\textrm{obs}} | \mathbf{X}^\mathsf{s}, \mathbf{w}^\mathsf{s}, \Phi, \Sigma, \mathbf{g}^\mathsf{s})$, which is useful in inferring parameters and hyperparameters of our proposed model. We then also obtain the posterior $\p(\mathbf{y}^\mathsf{s}_\textrm{mis}\,\big| \mathbf{y}^\mathsf{s}_\textrm{obs},\mathbf{X}^\mathsf{s}, \mathbf{w}^\mathsf{s}, \Phi, \Sigma, \mathbf{g})$ to estimate ITE and local ATE.

\subsubsection{The federated objective function}
\label{sec:objective-function}
Since estimating $\p(\mathbf{y}^\mathsf{s}_{\textrm{mis}}\,\big|\, \mathbf{y}^\mathsf{s}_{\textrm{obs}}, \mathbf{X}^\mathsf{s}, \mathbf{w}^\mathsf{s})$ exactly is intractable, we sidestep this intractability via a variational approximation \citep{kingma2013auto,blei2017variational}. To achieve this, we maximize the following evidence lower bound (ELBO) $\mathbf{L}$:
\begin{align}
    \log\p(\mathbf{y}_{\textrm{obs}}\,|\,\mathbf{X},\mathbf{w}) &= \log \int\p(\mathbf{y}_{\textrm{obs}},\mathbf{g}, \Phi, \Sigma\,|\,\mathbf{X},\mathbf{w}) d\mathbf{g}d\Phi d\Sigma\ge \sum_{\mathsf{s}=1}^m \mathbf{L}^\mathsf{s} =\vcentcolon\mathbf{L},\label{eq:loss}
\end{align}
where $\mathbf{L}^\mathsf{s} = \e_q[ \log\p(\mathbf{y}^\mathsf{s}_{\textrm{obs}} | \cdot)]  -\frac{1}{m}\sum_{z \in \{\mathbf{g}, \Phi, \Sigma\}}\D_{\text{KL}}(\q(z)\|\p(z))$. 
The conditional likelihood $\p(\mathbf{y}^\mathsf{s}_{\textrm{obs}} | \cdot)$ is obtained from Lemma~\ref{lem:3} by marginalizing out $\mathbf{y}^\mathsf{s}_{\textrm{mis}}$, i.e.,
\begin{align}
    \!\!\!\!\!\!\p(\mathbf{y}^\mathsf{s}_{\textrm{obs}} | \mathbf{X}^\mathsf{s}, \mathbf{w}^\mathsf{s}, \Phi, \Sigma, \mathbf{g}^\mathsf{s}) = \mathsf{N}(\mathbf{y}^\mathsf{s}_{\textrm{obs}};\mu_{\textrm{obs}}(\mathbf{X}^\mathsf{s}), \mathbf{K}_{\textrm{obs}}).
\end{align}
We observe that the above conditional likelihood is free of $\sigma_{21}$ and $\sigma_{12}$, which captures the correlation of two potential outcomes. Thus the posterior of these variables would coincide with their priors, i.e., the correlation cannot be learned but set as a prior. This is well-known as one of the potential outcome cannot be observed \citep{imbens2015causal}. 
In Eq.~(\ref{eq:loss}), the ELBO $\mathbf{L}$ is derived from the of joint marginal likelihood of all $m$ sources, and it is factorized into $m$ components $\mathbf{L}^\mathsf{s}$, each component corresponds to a source. This enables federated optimization of $\mathbf{L}$. The first term of $\mathbf{L}^\mathsf{s}$ is expectation of the conditional likelihood with respect to the variational posterior $q(\mathbf{g}, \Phi, \Sigma)$, thus this distribution is learned from data of all the sources. In the following, we present the factorization of this distribution. 

\textbf{Variational posterior distributions.} 
We apply the typical mean-field approximation to factorize among the variational posteriors $
\q(\Phi, \Sigma, \mathbf{g}) =\q(\Phi)\,\q(\Sigma)\,\q(\mathbf{g})$,
where\begin{align*}
    \q(\mathbf{g}) = \prod_{j\in\{0,1\}}\mathsf{N}(\mathbf{g}_j;h_j(\mathbf{\tilde{y}}_{\textrm{obs}}(0), \mathbf{\tilde{y}}_{\textrm{obs}}(1), \mathbf{\tilde{X}}, \mathbf{\tilde{w}}), \mathbf{U}),
\end{align*}
where we denote $\mathbf{\tilde{y}}_{\textrm{obs}}^\mathsf{s}(0)$, $\mathbf{\tilde{y}}_{\textrm{obs}}^\mathsf{s}(1)$, and $\tilde{w}^\mathsf{s}$ as the first four moments of the observed outcomes and treatment of the $\mathsf{s}$--th source, and $\mathbf{\tilde{X}} = [\mathbf{\tilde{x}}^{1},\!...,\mathbf{\tilde{x}}^{m}]^\top$, $\mathbf{\tilde{y}}(0) = [\tilde{y}_{\textrm{obs}}^{1}(0),\!...,\tilde{y}_{\textrm{obs}}^{m}(0)]^\top$, $\mathbf{\tilde{y}}(1) = [\tilde{y}_{\textrm{obs}}^{1}(1),\!...,\tilde{y}_{\textrm{obs}}^{m}(1)]^\top$, and $\mathbf{\tilde{w}} = [\tilde{w}^{1},\!...,\tilde{w}^{m}]^\top$,  $h_0(\cdot)$ and $h_1(\cdot)$ are the mean functions, $\mathbf{U}$ is the covariance matrix computed with a kernel function $\kappa(u^\mathsf{s}, u^{\mathsf{s}'})$, where $u^\mathsf{s}\vcentcolon= [\tilde{y}^\mathsf{s}_{\textrm{obs}}(0), \tilde{y}^\mathsf{s}_{\textrm{obs}}(1), \mathbf{\tilde{x}}^\mathsf{s}, \tilde{w}^\mathsf{s}]$. 
Since $\Phi$ and $\Sigma$ are positive semi-definite matrices, we model their variational posterior as Wishart distribution: $
	\q(\Phi) = \mathsf{Wishart}(\Phi;\mathbf{V}_q, d_q)$ and $\q(\Sigma) = \mathsf{Wishart}(\Sigma;\mathbf{S}_q, n_q)$,
	where $d_q, n_q$ are degrees of freedom and $\mathbf{V}_q, \mathbf{S}_q$ are the  positive semi-definite scale matrices. We set the form of these scale matrices as follows
\begin{align*}
\mathbf{V}_q = \begin{bmatrix}
\nu_{1}^2&\rho\nu_{1}\nu_2\\
\rho\nu_{1}\nu_2&\nu_{2}^2
\end{bmatrix}, \quad\mathbf{S}_q = \begin{bmatrix}
\delta_{1}^2&\eta\delta_{1}\delta_2\\
\eta\delta_{1}\delta_2&\delta_{2}^2
\end{bmatrix}.
\end{align*}
where $\nu_i, \rho, \delta_i, \eta$ are parameters to be learned and $\rho, \eta \in [0,1]$.

\textbf{Reparameterization.} To maximize the ELBO, we approximate the expectation in $\mathbf{L}^\mathsf{s}$ with Monte Carlo integration, which require drawing samples of $\mathbf{g}$, $\Phi$ and $\Sigma$ from their variabional distributions. This requires a reparameterization to allow the gradients to pass through the random variables $\mathbf{g}$, $\Phi$ and $\Sigma$. 
Since we model the correlation among each individual and the correlation between the two possible outcomes, the typical reparameterization of \citet{kingma2013auto} cannot be applied as this method only holds true with diagonal covariance matrix. The reparameterization trick we applied is more general
\begin{align*}
&\mathbf{g}_j = h_j(\mathbf{\tilde{y}}_{\textrm{obs}}(0), \mathbf{\tilde{y}}_{\textrm{obs}}(1), \mathbf{\tilde{X}}, \mathbf{\tilde{w}}) + \mathbf{U}^{\frac{1}{2}} \bm{\xi}_j,\quad j\in\{0,1\},
\end{align*}
where $\bm{\xi}_j \sim \mathsf{N}(\bm{0},\mathbf{I}_m)$ and $\mathbf{U}^{\frac{1}{2}}$ is the Cholesky decomposition matrix of $\mathbf{U}$.
Since $\q(\Phi)$ is modeled as Wishart distribution, we introduce the following procedure to draw $\Phi$:
	\begin{align*}
	\Phi = \mathbf{V}_q^{\frac{1}{2}} \bm{\zeta} ( \mathbf{V}_q^{\frac{1}{2}})^\top, \quad\bm{\zeta} \sim \mathsf{Wishart}(\mathbf{I}_2, d_q),
	\end{align*}
where $\mathbf{V}_q^{\frac{1}{2}}$ is the Choleskey decomposition matrix of $\mathbf{V}_q$. Likewise, we also apply this procedure to draw $\Sigma$.

\textbf{The Federated optimization algorithm.} 
With the above designed model and its objective function, we can compute gradients of the learnable parameters separately in each source without sharing data to a central server. Thus, it satisfies the privacy constraints. We summarize our inference procedure in Algorithm~\ref{algo:maximize-elbo}. \begin{center}
{\IncMargin{1.3em}
\begin{algorithm}
\setstretch{0}
\caption{FedCI: Federated causal inference}
\small
\label{algo:maximize-elbo}
\SetKwInOut{Parameter}{Parameters}
\Parameter{Let $\Theta$ be set of parameters}
\Begin{
 		Initialize $\Theta$ and send to all source machines\;
	    \Repeat{stopping condition}{
    		\For{each source machine $\mathsf{s} \in \{1,2,\dots,m\}$}{
    			Compute $\nabla_\Theta \mathbf{L}^\mathsf{s}$ and send to server\;
    		}
    		In the central server, do the following steps:\\
    		\Begin{
    		Collect gradients from all sources and compute $\nabla_\Theta\mathbf{L} = \sum_{\mathsf{s}=1}^m \nabla_\Theta\mathbf{L}^\mathsf{s}$\;
    		Update $\Theta \leftarrow \Theta + \mathsf{learning\_rate} \times\nabla_\Theta\mathbf{L} $\;
    		Broadcast the new $\Theta$ to all sources\;
    		}
		}
}
\end{algorithm}
\DecMargin{1.3em}
\vskip -16pt
}
\end{center}
\subsubsection{How data from all sources help prediction of causal effects in a specific source?}
Remember that the key to estimate ITE and ATE is to find the predictive distribution $\p(\mathbf{y}_\textrm{mis}\,\big|\, \mathbf{y}_\textrm{obs}, \mathbf{X}, \mathbf{w})$. This distribution can be estimated by
the following relation: \begin{align*}
    &\p(\mathbf{y}_\textrm{mis}\big|\mathbf{y}_\textrm{obs}, \mathbf{X}, \mathbf{w}) \simeq \e_{\q} \bigg[\prod_{\mathsf{s}=1}^m\p(\mathbf{y}^\mathsf{s}_\textrm{mis}\big| \mathbf{y}^\mathsf{s}_\textrm{obs},\mathbf{X}^\mathsf{s}, \mathbf{w}^\mathsf{s}, \Phi, \Sigma, \mathbf{g})\bigg],
\end{align*}
where the expectation is with respect to the variational distribution $\q(\Phi,\Sigma,\mathbf{g})$, and
\begin{align*}
    &\p(\mathbf{y}^\mathsf{s}_\textrm{mis}\,\big| \mathbf{y}^\mathsf{s}_\textrm{obs},\mathbf{X}^\mathsf{s}, \mathbf{w}^\mathsf{s}, \Phi, \Sigma, \mathbf{g}) = \mathsf{N}\left(\mathbf{y}^\mathsf{s}_\textrm{mis};\mathbf{m}_{\textrm{mo}}^\mathsf{s}, \mathbf{S}_{\textrm{mo}}^\mathsf{s}\right),\\
    &\mathbf{m}_{\textrm{mo}}^\mathsf{s} = \mu_{\textrm{mis}}(\mathbf{X}^\mathsf{s}) + (\mathbf{K}_{\textrm{om}}^\mathsf{s})^\top(\mathbf{K}_{\textrm{obs}}^\mathsf{s})^{-1}(\mathbf{y}^\mathsf{s}_{\textrm{obs}} - \mu_{\textrm{obs}}(\mathbf{X}^\mathsf{s})),\\
    &\mathbf{S}_{\textrm{mo}}^\mathsf{s} = \mathbf{K}_{\textrm{mis}}^\mathsf{s} - (\mathbf{K}_{\textrm{om}}^\mathsf{s})^\top(\mathbf{K}_{\textrm{obs}}^\mathsf{s})^{-1}\mathbf{K}_{\textrm{om}}^\mathsf{s}.
\end{align*}
To understand why data from all the sources can help predict causal effects in a source $\mathsf{s}$, we observe that \begin{align}
    \p(\mathbf{y}^\mathsf{s}_\textrm{mis}\,\big| \mathbf{y}_\textrm{obs}, \mathbf{X}, \mathbf{w}) &\simeq \e_{\q} \big[p(\mathbf{y}^\mathsf{s}_\textrm{mis}\big| \mathbf{y}^\mathsf{s}_\textrm{obs},\mathbf{X}^\mathsf{s}, \mathbf{w}^\mathsf{s}, \Phi, \Sigma, \mathbf{g})\big]\nonumber\\
&= p(\mathbf{y}^\mathsf{s}_\text{mis}\,\big| \underbrace{\addstackgap[1.3pt]{\textcolor{red!80!black}{$\mathbf{y}^\mathsf{s}_\text{obs}, \mathbf{X}^\mathsf{s}, \mathbf{w}^\mathsf{s}$}}}_\textbf{(i)}, \underbrace{\addstackgap[4.6pt]{$ \textcolor{green!60!black}{\Theta} $}}_\textbf{(ii)}, \underbrace{\addstackgap[2pt]{\textcolor{blue!80!black}{$\mathbf{\tilde{y}}_{\textrm{obs}}(0), \mathbf{\tilde{y}}_{\textrm{obs}}(1), \mathbf{\tilde{X}}, \mathbf{\tilde{w}}$}}}_\textbf{(iii)}),\label{eq:predictive-dist}
\end{align}
Eq.~(\ref{eq:predictive-dist}) is an approximation of the predictive distribution of the missing outcomes $\mathbf{y}^\mathsf{s}_\textrm{mis}$ and it depends on the following three components:
\begin{enumerate}[noitemsep,topsep=0pt,leftmargin=*,label=\textbf{(\roman*).}]
    \item The observed outcomes, covariates and treatment assignments from the same source $\mathsf{s}$, and
    \item The shared parameters $\Theta$ learned from data of all the sources, and
    \item Sufficient statistics of the observed data from all the sources.\end{enumerate}
The two last components \textbf{(ii)} and \textbf{(iii)} indicate that the predictive distribution in source $\mathsf{s}$ utilized knowledge from all the sources through $\Theta$ and the sufficient statistics $[\mathbf{\tilde{y}}_{\textrm{obs}}(0), \mathbf{\tilde{y}}_{\textrm{obs}}(1), \mathbf{\tilde{X}}, \mathbf{\tilde{w}}]$. 
This explain why data from all the sources help predict missing outcomes in source $\mathsf{s}$.

\section{Experiments}
\label{sec:experiment}

\textbf{Baselines and the aims of our experiments.} 
In this section, we first perform experiments to examine the performance of FedCI. We then compare the performance of FedCI against recent findings, such as BART \citep{hill2011bayesian}, 
CEVAE \citep{louizos2017causal}, 
OrthoRF \citep{oprescu2019orthogonal}, 
X-learner \citep{kunzel2019metalearners}, and R-learner \citep{nie2020quasi}. Note that all these work do not consider causal effects in a federated setting. The aim of this analysis is to show the efficacy of our method compared to the baselines trained in three different cases: (\textbf{1}) training a local model on each source data, (\textbf{2}) training a global model with the combined data of all sources, (\textbf{3}) using bootstrap aggregating (also known as bagging; is an ensemble learning method) of \citet{breiman1996bagging} where $m$ models are trained separately on each source data and then averaging the predicted treatment effects of each model. Note that case (\textbf{2})  \textit{violates} individuals' privacy rights and is only used for comparison purposes. In general, we expect that the performance of FedCI is close to that of the performance of the baselines in case (\textbf{2}).
Implementation of CEVAE is readily available \citep{louizos2017causal}. For the implementation of BART \citep{hill2011bayesian}, we use the package \texttt{BartPy}, which is also available online. For X-learner \citep{kunzel2019metalearners} and R-learner \citep{nie2020quasi}, we use the package \texttt{causalml} \citep{chen2020causalml}. In both methods, we use \texttt{xgboost.XGBRegressor} as learners for the outcomes. 
For OrthoRF \citep{oprescu2019orthogonal}, we use the package \texttt{econml} \citep*{econml}. For all the methods, we fine-tune the learning rate in $\{10^{-1}, 10^{-2}, 10^{-3}, 10^{-4}\}$ and regularizers in $\{10^{1}, 10^{0}, 10^{-1}, 10^{-2}, 10^{-3}\}$.

\textbf{Evaluation metrics.} We report two evaluation metrics: (i) precision in estimation of heterogeneous effects (PEHE) \citep{hill2011bayesian}: $\epsilon_\textrm{PEHE} \vcentcolon= \sum_{s=1}^m\sum_{i=1}^{n_\mathsf{s}}(\tau^{\mathsf{s}}_i - \hat{\tau}^{\mathsf{s}}_i)^2/(m n_\mathsf{s})$ for evaluating ITE, 
and (ii) absolute error: $\epsilon_\textrm{ATE} \vcentcolon=  |\tau-\hat{\tau}|$ for evaluating ATE, where $\tau_{i}^{\mathsf{s}}$ and $\tau$ are the \textit{true} ITE and \textit{true} ATE, respectively, and $\hat{\tau}_{i}^{\mathsf{s}}$,  $\hat{\tau}$ are their estimates. Note that these evaluation metrics are for point estimates of the treatment effects. In our case, the point estimates are the mean of ITE and ATE in their predictive distributions.

\subsection{Synthetic data}
\label{sec:synthetic-data}
\textbf{Data description.} Obtaining ground truth for evaluating causal inference algorithm is a challenging task. Thus, most of the state-of-the-art methods are evaluated using synthetic or semi-synthetic datasets. In this experiment, the synthetic data is simulated with the following distributions:
\begin{align*}
    x_{ij} &\sim \mathsf{U}[-1,1], &y_i(0) &\sim \mathsf{N}(\lambda(b_0 + \mathbf{x}_i^\top \mathbf{b}_1), \sigma_0^2), \\
    w_i &\sim \mathsf{Bern}(\varphi(a_0 + \mathbf{x}_i^\top \mathbf{a}_1)), &y_i(1) &\sim \mathsf{N}(\lambda(c_0 + \mathbf{x}_i^\top \mathbf{c}_1), \sigma_1^2),
\end{align*}
where $\varphi(\cdot)$ denotes the sigmoid function, $\lambda(\cdot)$ denotes the softplus function, and $\mathbf{x}_i = [x_{i1},\!...,x_{id_x}]^\top \in \mathbb{R}^{d_x}$ with $d_x=20$. We simulate two synthetic datasets: DATA-1 and DATA-2. For DATA-1, the ground truth parameters are randomly set as follows: $\sigma_0=\sigma_1=1$,  $(a_0,b_0,c_0)=(0.6,0.9,2.0)$, $\mathbf{a}_1 \sim \mathsf{N}(\mathbf{0}, 2\cdot\mathbf{I}_{d_x})$, $\mathbf{b}_1 \sim \mathsf{N}(\mathbf{0}, 2\cdot\mathbf{I}_{d_x})$, $\mathbf{c}_1 \sim \mathsf{N}(\mathbf{1}, 2\cdot\mathbf{I}_{d_x})$, where $\mathbf{1}$ is a vector of ones and $\mathbf{I}_{d_x}$ is an identity matrix. For DATA-2, we set $(b_0,c_0) = (6,30)$, $\mathbf{b}_1 \sim \mathsf{N}(10\cdot\mathbf{1}, 2\cdot\mathbf{I}_{d_x})$, $\mathbf{c}_1 \sim \mathsf{N}(15\cdot\mathbf{1}, 2\cdot\mathbf{I}_{d_x})$, and the other parameters are set similar to that of DATA-1. The purpose is to make two different scales of the outcomes for the two datasets. For each dataset, we simulate $10$ replications with $n = 5000$ records. We only keep $\{(y_i, w_i, \mathbf{x}_i)\}_{i=1}^n$ as the observed data, where $y_i = y_i(0)$ if $w_i=0$ and $y_i = y_i(1)$ if $w_i=1$. We divide the data into five sources, each consists of $n_\mathsf{s}=1000$ records. In each source, we use $50$ records for training, $450$ for testing and $400$ for validation. In the following, we report the evaluation metrics and their standard errors over the 10 replications.

\begin{figure}\centering
    \includegraphics[width=0.65\textwidth]{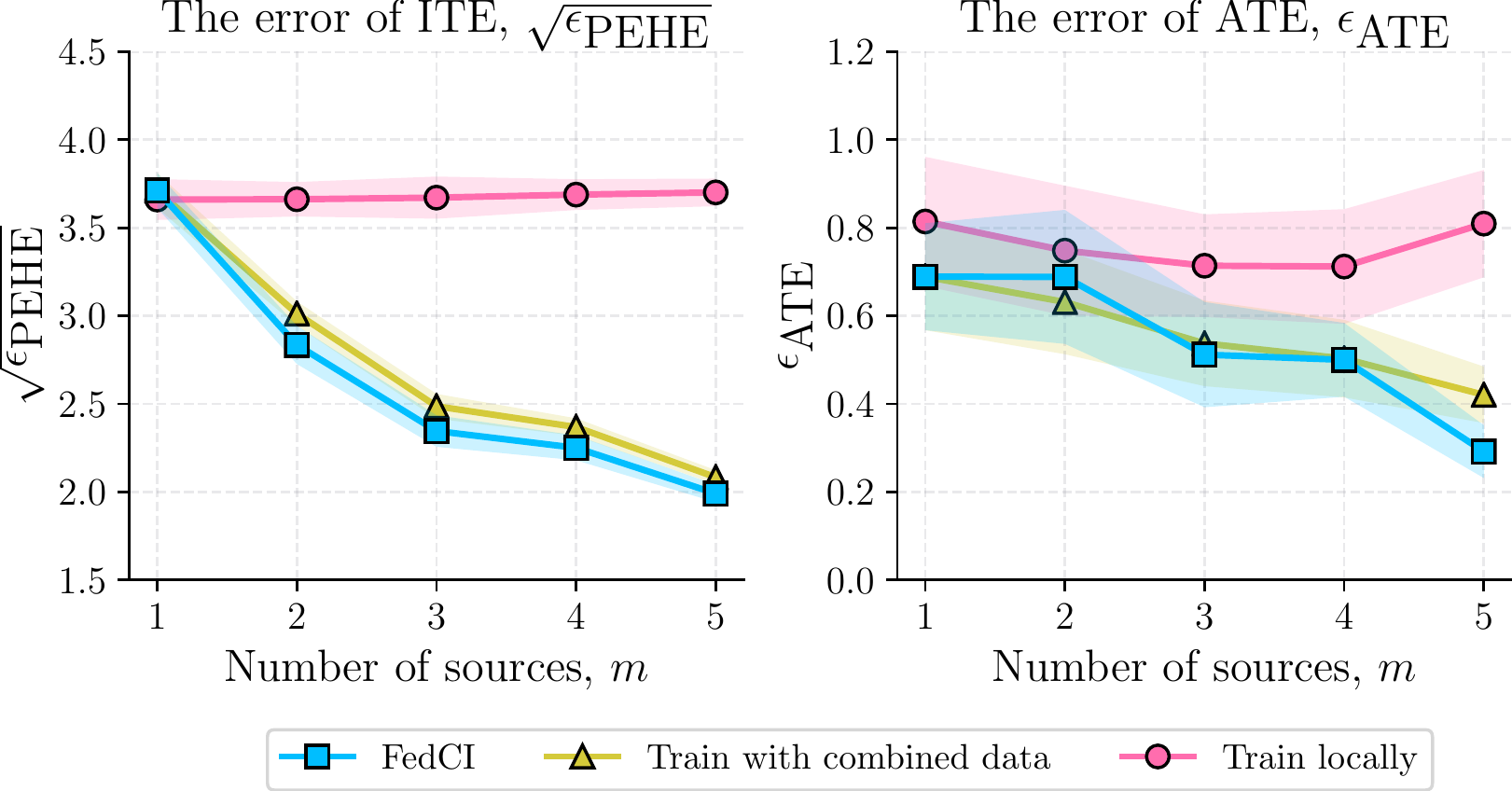}
\caption{Analysis on DATA-1.}
\label{fig:fedci-alalysis}
\end{figure}
\begin{figure}
        \centering
    \includegraphics[width=0.65\textwidth]{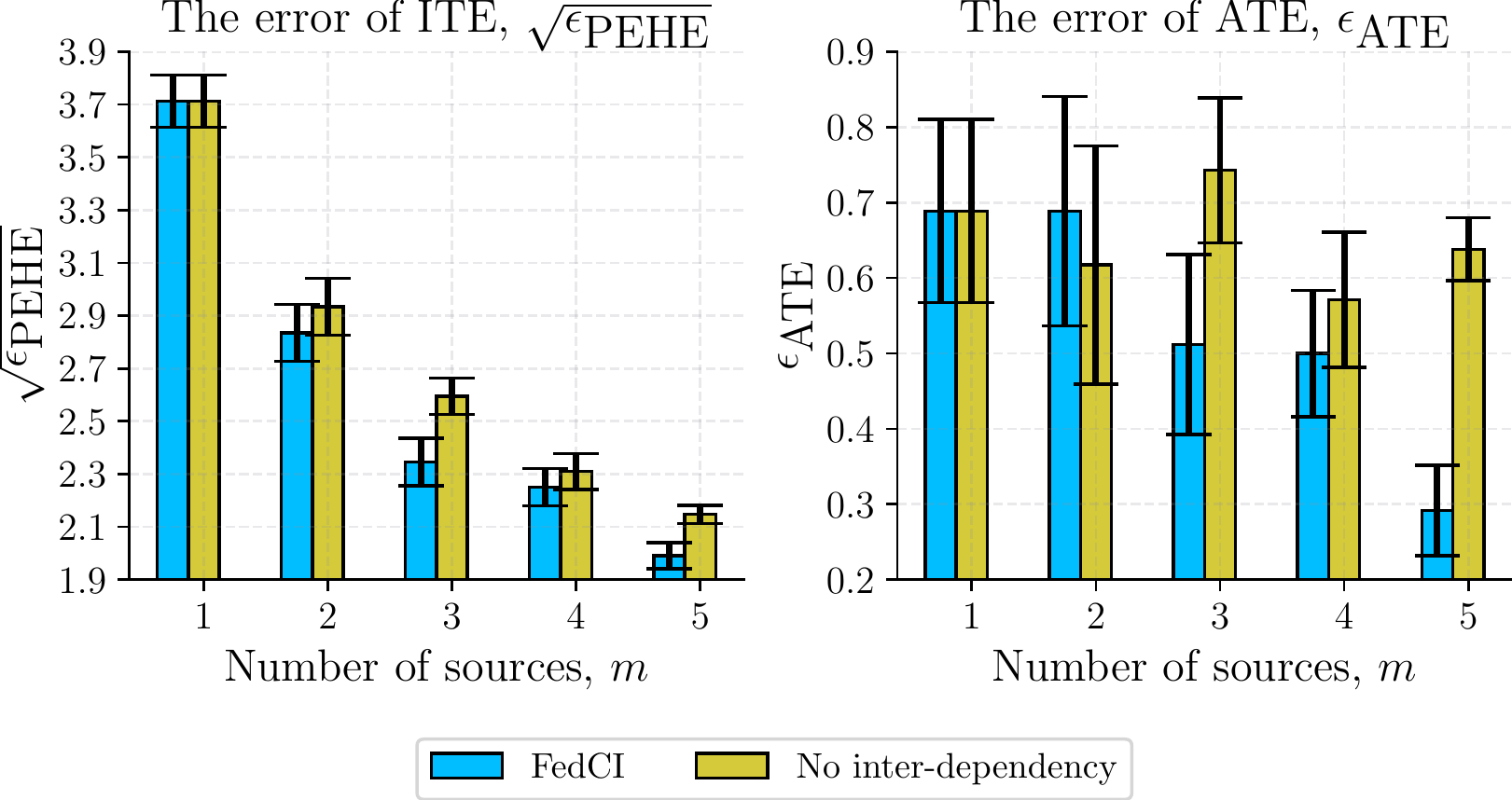}
\caption{The impact of inter-dependency on DATA-1.}
\label{fig:fedci-inter-dependency-alalysis}
\end{figure}

The above parameters chosen for this simulation study satisfy Assumption~\ref{assumption:ignorability} since $y_i(0)$ and $y_i(1)$ are independent with $w_i$ given $\mathbf{x}_i$. Assumption~\ref{assumption:sutva} is respected as the treatment treatment on an individual $i$ does not effect the outcome of another individual $j$ ($i\neq j$). Since we fixed the dimension of $\mathbf{x}_i$ and draw it from the same distribution, Assumption~\ref{assumption:share-covariates} is implicitly satisfied. It is important to note that Assumption~\ref{assumption:unique-ident} and the preprocessing procedure are not necessary since each record that is drawn from the above distributions is attributed to one individual. This necessarily means that there are no duplicates of individuals in more than one source. In a real life setting, in the case when there are individuals appearing in multiple sources, Assumption~\ref{assumption:unique-ident} needs to hold, and the preprocessing procedure described in Section~\ref{sec:assumptions} has to be performed to exclude those repeated individuals from the training process. 

\textbf{FedCI is as good as training on combined data.} Figure~\ref{fig:fedci-alalysis} reports the three evaluation metrics of FedCI compared with two baselines: training on combined data and training locally on each data source. As expected, the figures show that the errors of FedCI are as low as that of training on the combined data. This result verifies the efficacy of the proposed federated algorithm.

\textbf{Inter-dependency component analysis.} We study the impact of the inter-dependency component (see Section~\ref{sec:model}) by removing it from the model. Figure~\ref{fig:fedci-inter-dependency-alalysis} presents the errors of FedCI compared with `no inter-dependency' (FedCI without inter-dependency). The figures show that the errors in predicting ITE and ATE of `no inter-dependency' seems to be higher than those of FedCI. This result showcases the importance of our proposed inter-dependency component.

\textbf{Contrasting with existing baselines.} In this experiment, we compare FedCI with the existing baselines. Note that all the baselines do not consider estimating causal effects on multiple sources with privacy constraints. Thus, we train them in three cases as explained earlier: \textbf{(1)} train locally ($\mathsf{loc}$), \textbf{(2)} train with combined data ($\mathsf{com}$), and \textbf{(3)} train with bootstrap aggregating ($\mathsf{agg}$). Note that case (\textbf{2}) violates privacy constraints. In general, we expect that the error of FedCI to be close to case \textbf{(2)} of the baselines. Table~\ref{tab:error-synthetic}~and~\ref{tab:error-synthetic-2} reports the  performance of each method in estimating ATE and ITE. Regardless of different scales on the two synthetic dataset, the figure shows that FedCI achieves competitive results compared to all the baselines. In particular, FedCI is among the top-3 performances among all the methods. Importantly, FedCI obtains lower errors than those of BART$_\mathsf{com}$, X-Learner$_\mathsf{com}$, R-Learner$_\mathsf{com}$, and OthoRF$_\mathsf{com}$, which were trained on combined data and thus violate privacy constraints. Compare with CEVAE$_\mathsf{com}$, FedCI is better than this method in predicting ITE and comparable with this method in predicting ATE (slightly higher errors). However, we emphasize again that this result is expected since we proposed a federated learning algorithm while CEVAE$_\mathsf{com}$ is not a federated one.

\begin{figure}
\centering
    \includegraphics[width=0.75\textwidth]{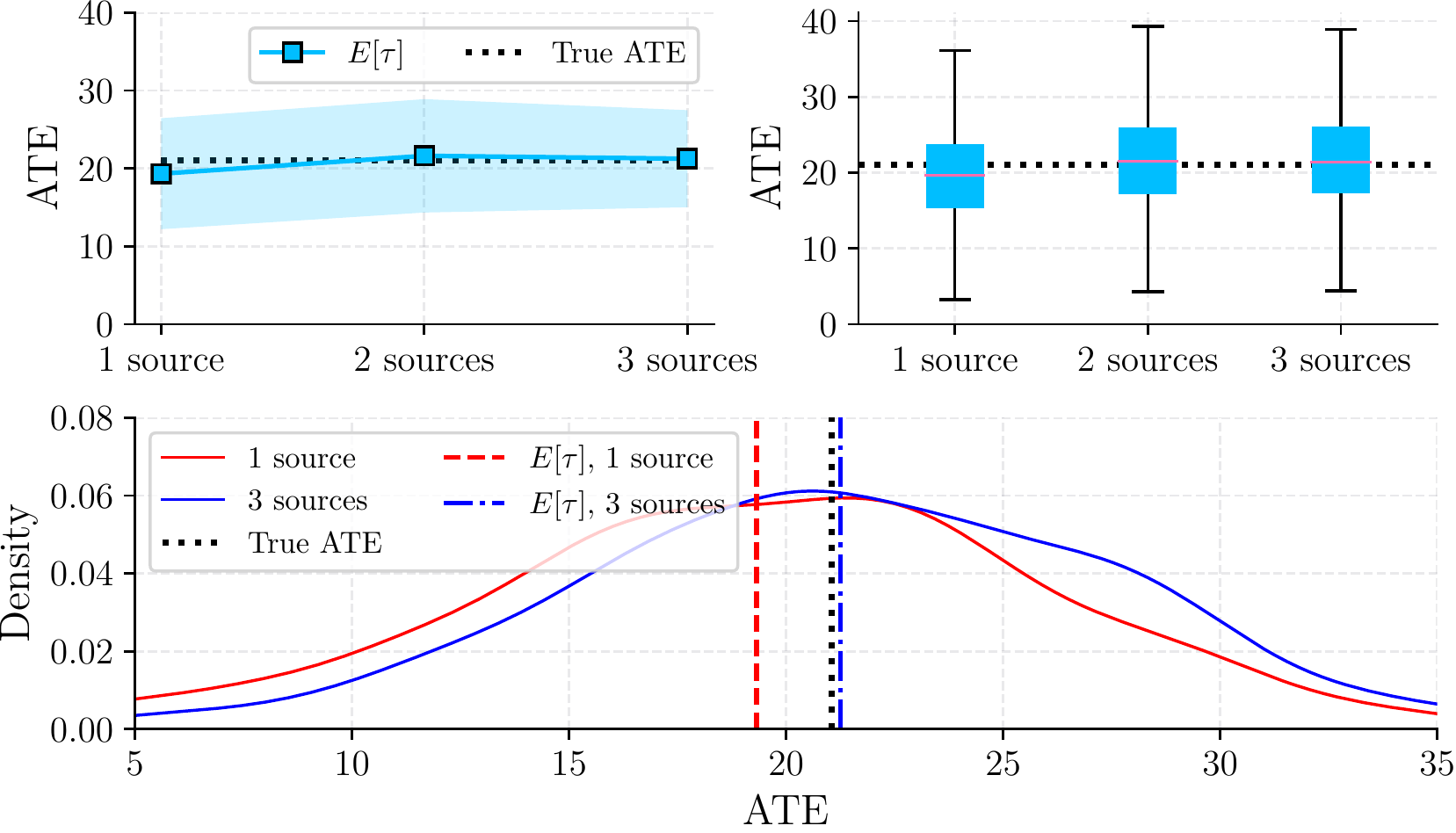}
    \vspace{-6pt}
\caption{The estimated distribution of ATE on source \#1 of DATA-2. The dotted black lines represent the true ATE.}
\label{fig:fedci-uncertainty-analysis-synthetic}
\end{figure}

\textbf{The estimated distribution of ATE.} To analyse uncertainty, we present in  Figure~\ref{fig:fedci-uncertainty-analysis-synthetic} the estimated distribution of ATE in the first source ($\mathsf{s}=1$). The figures show that the true ATE is covered by the estimated interval and the estimated mean ATE shifts towards its true value (dotted lines) when more data sources are used. This result might give helpful information for user.

\begin{table}
\centering
\caption{Out-of-sample errors on DATA-1 where top-3 performances are highlighted in bold (lower is better). The dashes (---) in `$\mathsf{loc}$' and `$\mathsf{agg}$' indicate that the numbers are the same as those of `$\mathsf{com}$'.}
\vspace{3pt}
\label{tab:error-synthetic}
\setlength{\tabcolsep}{2.7pt}
\footnotesize
\begin{tabular}{@{}lcccccc@{}}
\toprule
\multirow{2}{*}{Method}                                           & \multicolumn{3}{c}{The error of ITE ($\sqrt{\epsilon_\text{PEHE}}$)} & \multicolumn{3}{c}{The error of ATE ( $\epsilon_\text{ATE}$)} \\ \cmidrule(lr){2-4}\cmidrule(lr){5-7} 
                                                                  & 1 source      & 3 sources     & 5 sources     & 1 source      & 3 sources     & 5 sources    \\ \cmidrule(r){1-1}\cmidrule(lr){2-4}\cmidrule(lr){5-7}
BART$_\mathsf{loc}$                                                       & ---    & 6.04$\pm$.05    & 6.02$\pm$.04    & ---    & 0.59$\pm$.14    & 0.53$\pm$.10   \\
X-Learner$_\mathsf{loc}$                                                  & ---    & 5.81$\pm$.13    & 5.77$\pm$.09    & ---    & 0.44$\pm$.24    & 0.51$\pm$.13   \\
R-Learner$_\mathsf{loc}$                                                  & ---    & 5.94$\pm$.05    & 5.94$\pm$.03    & ---    & 0.65$\pm$.05    & 0.66$\pm$.02   \\
OthoRF$_\mathsf{loc}$                                                     & ---   & 5.83$\pm$.12    & 6.23$\pm$.13    & ---    & \textbf{0.31$\pm$.08}    & 0.52$\pm$.10   \\
CEVAE$_\mathsf{loc}$                                                     & ---   & 3.82$\pm$.09    & 3.50$\pm$.06    & ---    & 0.63$\pm$.11    & 0.52$\pm$.03   \\\cmidrule(r){1-1}\cmidrule(lr){2-4}\cmidrule(lr){5-7}
BART$_\mathsf{agg}$                                                 & ---             & 5.97$\pm$.05              & 5.94$\pm$.03              & ---             & 0.64$\pm$.14              & 0.47$\pm$.11             \\
X-Learner$_\mathsf{agg}$ & ---             & 5.18$\pm$.09    & 5.09$\pm$.05    & ---             & 0.46$\pm$.24    & 0.52$\pm$.13   \\
R-Learner$_\mathsf{agg}$ & ---             & 5.94$\pm$.05    & 5.93$\pm$.03    & ---             & 0.65$\pm$.05    & 0.66$\pm$.03   \\
OthoRF$_\mathsf{agg}$    & ---             & 4.19$\pm$.13    & 3.66$\pm$.08              & ---             & 0.36$\pm$.13    & 0.48$\pm$.12             \\
CEVAE$_\mathsf{agg}$                                                     & ---    & 3.65$\pm$.10    & 2.99$\pm$.06    & ---    & 0.41$\pm$.05    & 0.37$\pm$.04   \\\cmidrule(r){1-1}\cmidrule(lr){2-4}\cmidrule(l){5-7}
BART$_\mathsf{com}$                                                     & 5.98$\pm$.06             & 5.97$\pm$.06    & 5.93$\pm$.03    & 0.83$\pm$.11             & 0.56$\pm$.16    & 0.38$\pm$.09   \\
X-Learner$_\mathsf{com}$    & \textbf{5.48$\pm$.15}             & 4.60$\pm$.09    & 4.15$\pm$.04    & 0.93$\pm$.22             & 0.60$\pm$.11    & \textbf{0.30$\pm$.07}   \\
R-Learner$_\mathsf{com}$     & 5.93$\pm$.06             & 5.73$\pm$.08    & 5.54$\pm$.06    & 0.78$\pm$.10             & 0.47$\pm$.09    & \textbf{0.30$\pm$.07}   \\
OthoRF$_\mathsf{com}$        & 5.86$\pm$.40             & \textbf{3.60$\pm$.12}    & \textbf{2.94$\pm$.05}    & \textbf{0.55$\pm$.14}             & 0.45$\pm$.14    & 0.34$\pm$.09   \\
CEVAE$_\mathsf{com}$        & \textbf{3.79$\pm$.07}             & \textbf{2.85$\pm$.06}    & \textbf{2.72$\pm$.04}    & \textbf{0.51$\pm$.13}             & \textbf{0.23$\pm$.07}    & \textbf{0.20$\pm$.06}   \\\cmidrule(r){1-1}\cmidrule(lr){2-4}\cmidrule(lr){5-7}
FedCI                                                             & \textbf{3.71$\pm$.10}     & \textbf{2.35$\pm$.09}    & \textbf{1.99$\pm$.05}    & \textbf{0.69$\pm$.12}    & \textbf{0.31$\pm$.12}    & \textbf{0.29$\pm$.06}   \\ \bottomrule
\end{tabular}
\vskip -6pt
\end{table}

\begin{table}
\centering
\caption{Out-of-sample errors on DATA-2 where top-3 performances are highlighted in bold (lower is better). Please see the full table in Appendix, which includes `$\mathsf{loc}$' \& `$\mathsf{agg}$'.}
\vspace{3pt}
\label{tab:error-synthetic-2}
\setlength{\tabcolsep}{2.7pt}
\footnotesize
\begin{tabular}{@{}lcccccc@{}}
\toprule
\multirow{2}{*}{Method}                                           & \multicolumn{3}{c}{The error of ITE ($\sqrt{\epsilon_\text{PEHE}}$)} & \multicolumn{3}{c}{The error of ATE ( $\epsilon_\text{ATE}$)} \\ \cmidrule(lr){2-4}\cmidrule(lr){5-7} 
                                                                  & 1 source      & 3 sources     & 5 sources     & 1 source      & 3 sources     & 5 sources    \\ \cmidrule(r){1-1}\cmidrule(lr){2-4}\cmidrule(lr){5-7}
BART$_\mathsf{com}$                                                     & \textbf{18.0$\pm$0.4}             & \textbf{17.7$\pm$0.2}    & 17.4$\pm$0.1    & \textbf{3.54$\pm$1.3}             & 2.94$\pm$0.8    & \textbf{1.84$\pm$0.5}   \\
X-Learner$_\mathsf{com}$    & 21.1$\pm$0.9             & 17.9$\pm$0.4    & \textbf{16.2$\pm$0.2}    & 4.55$\pm$1.4             & 3.29$\pm$1.0    & 2.37$\pm$0.8   \\
R-Learner$_\mathsf{com}$     & 25.9$\pm$0.6             & 23.5$\pm$0.5    & 21.3$\pm$0.4    & 19.0$\pm$0.8             & 15.6$\pm$0.7    & 12.3$\pm$0.6   \\
OthoRF$_\mathsf{com}$        & 37.8$\pm$2.7             & \textbf{10.7$\pm$0.5}    & \textbf{9.83$\pm$0.5}    & 7.88$\pm$2.2             & \textbf{1.99$\pm$0.4}    & 2.36$\pm$0.6   \\
CEVAE$_\mathsf{com}$        & \textbf{20.1$\pm$0.5}             & 18.4$\pm$0.6    & 16.6$\pm$0.6    & \textbf{1.50$\pm$0.3}             & \textbf{1.38$\pm$0.4}    & \textbf{1.89$\pm$0.2}   \\\cmidrule(r){1-1}\cmidrule(lr){2-4}\cmidrule(lr){5-7}
FedCI                                                             & \textbf{9.28$\pm$0.4}     & \textbf{6.34$\pm$0.2}    & \textbf{5.53$\pm$0.1}    & \textbf{2.37$\pm$0.5}    & \textbf{1.47$\pm$0.4}    & \textbf{0.74$\pm$0.2}   \\ \bottomrule
\end{tabular}
\end{table}

\begin{table}
\centering
\caption{Out-of-sample errors on IHDP dataset where top-3 performances are highlighted in bold (lower is better). The dashes (---) in `$\mathsf{agg}$' indicate that the numbers are the same as those of `$\mathsf{com}$'. Please see the full table in Appendix.}
\vspace{3pt}
\label{tab:error-ihdp}
\setlength{\tabcolsep}{2.7pt}
\footnotesize
\begin{tabular}{@{}lcccccc@{}}
\toprule
\multirow{2}{*}{Method}                                           & \multicolumn{3}{c}{The error of ITE ($\sqrt{\epsilon_\text{PEHE}}$)} & \multicolumn{3}{c}{The error of ATE ( $\epsilon_\text{ATE}$)} \\ \cmidrule(lr){2-4}\cmidrule(lr){5-7} 
                                                                  & 1 source      & 2 sources     & 3 sources     & 1 source      & 2 sources     & 3 sources    \\ \cmidrule(r){1-1}\cmidrule(lr){2-4}\cmidrule(lr){5-7}
BART$_\mathsf{agg}$                                                 & ---             & 4.05$\pm$1.9              & \textbf{3.69$\pm$1.8}              & ---             & 2.09$\pm$1.0              & 1.30$\pm$0.5             \\
X-Learner$_\mathsf{agg}$ & ---             & 3.98$\pm$1.5    & 4.28$\pm$1.9    & ---             & 1.51$\pm$0.7    & \textbf{0.83$\pm$0.5}   \\
R-Learner$_\mathsf{agg}$ & ---             & 4.76$\pm$1.3    & 4.46$\pm$1.6    & ---             & 1.92$\pm$0.5    & 1.41$\pm$0.2   \\
OthoRF$_\mathsf{agg}$    & ---             & \textbf{3.40$\pm$1.1}    & 4.26$\pm$1.9              & ---             & \textbf{0.87$\pm$0.3}    & 1.20$\pm$0.6             \\
CEVAE$_\mathsf{agg}$    & ---             & 3.63$\pm$0.7    & 3.73$\pm$0.5              & ---             & \textbf{0.92$\pm$0.2}    & 0.84$\pm$0.5             \\\cmidrule(r){1-1}\cmidrule(lr){2-4}\cmidrule(l){5-7}
BART$_\mathsf{com}$                                                 & 5.98$\pm$2.7             & 4.32$\pm$2.1              & 4.04$\pm$2.0              & \textbf{1.80$\pm$1.1}             & 2.09$\pm$1.1              & 1.21$\pm$0.6             \\
X-Learner$_\mathsf{com}$ & \textbf{4.22$\pm$1.6}             & 4.15$\pm$1.5    & 4.06$\pm$1.8    & \textbf{1.64$\pm$0.7}             & 1.93$\pm$0.8    & 0.84$\pm$0.4   \\
R-Learner$_\mathsf{com}$ & 6.97$\pm$2.1             & 4.43$\pm$1.4    & 4.47$\pm$1.7    & 3.15$\pm$0.5             & 1.34$\pm$0.5    & 1.10$\pm$0.3   \\
OthoRF$_\mathsf{com}$    & 4.49$\pm$1.9             & 3.81$\pm$1.3    & 3.75$\pm$1.5              & 1.86$\pm$0.8             & 1.61$\pm$0.6    & 1.56$\pm$0.8\\ 
CEVAE$_\mathsf{com}$        & \textbf{3.16$\pm$0.6}             & \textbf{2.34$\pm$0.6}    & \textbf{2.31$\pm$0.7}    & 2.02$\pm$0.4             & \textbf{0.53$\pm$0.1}    & \textbf{0.48$\pm$0.2}   \\\cmidrule(r){1-1}\cmidrule(lr){2-4}\cmidrule(lr){5-7}
FedCI                                                             & \textbf{2.88$\pm$0.8}     & \textbf{2.36$\pm$0.5}    & \textbf{2.35$\pm$0.6}    & \textbf{1.43$\pm$0.7}    & 1.03$\pm$0.4    & \textbf{0.51$\pm$0.2}   \\ \bottomrule
\end{tabular}
\end{table}

\subsection{IHDP data}
\label{sec:ihdp}
\textbf{Data description.} The Infant Health and Development Program (IHDP) \citep{hill2011bayesian} is a randomized study on the impact of specialist visits (the treatment) on the cognitive development of children (the outcome). The dataset consists of 747 records with 25 covariates describing properties of the children and their mothers. 
The treatment group includes children who received specialist visits and control group includes children who did not receive. 
For each unit, a treated and a control outcome are simulated using the numerical schemes provided in the NPCI package \citep{dorie2016npci}, thus allowing us to know the \textit{true} individual treatment effect.
We use 10 replicates of the dataset in this experiment. For each replicate, we  divide into three sources, each consists of 249 records. For each source, we then split it into three equal sets for the purpose of training, testing, and validating the models. We report the mean and standard error of the evaluation metrics over 10 replicates of the data. This dataset satisfies the Assumptions~\ref{assumption:ignorability},~\ref{assumption:sutva},~\ref{assumption:share-covariates}. Assumption~\ref{assumption:unique-ident} is redundant since there is are no repetitions of individuals in this dataset.

\textbf{Results and discussion.} Similar to the experiment for synthetic dataset,  here we also train the baselines in three cases as explained earlier. We also expect that the errors of FedCI to be close to the baseline trained with combined data ($\mathsf{com}$). The result reported in Table~\ref{tab:error-ihdp} shows that the FedCI achieves a competitive results compared to the baselines (we skipped the first case ($\mathsf{loc}$), please see Appendix for the full table). Indeed, FedCI is in the top-3 performances among all the methods. This result again verifies that FedCI can be used to estimate causal effects effectively under some privacy constraints of the data sources. The estimated distribution of ATE is presented in Appendix due to limited space.

\section{Conclusion}
\label{sec:conclusion}
We introduced a causal inference paradigm via a reformulation of multi-output GPs to learn causal effects, while keeping private data at their local sites. A modular inference method whose ELBO can be decomposed additively across data sources is presented. We posit that our formulation would prove useful in a diverse range of use cases within a causal inference setting on different range of applications.

We note that the inherently use of GP in our approach would in fact incur the computational time of inverse covariance matrix in each source of cubic time complexity. 
A possible future work direction is to reformulate this in terms of the recent sparse Gaussian Process models.%

\bibliographystyle{apalike}
\bibliography{ref}

\appendix

\vspace{2cm}
\begin{center}
    \centering \LARGE\bfseries Appendix: \\Federated Estimation of Causal Effects\\from Observational Data
\end{center}
\vspace{1.5cm}
\section{Additional experimental results}
\subsection{IHDP dataset}
In this section, we present additional  experimental results on the IHDP dataset. The results here were not presented in the main text due to limited space. In Table~\ref{tab:error-ihdp-appendix} (five first rows), we present additional results of the baselines trained locally ($\mathsf{loc}$). Similar to the experiments on synthetic data, the results here show that FedCI achieves much smaller errors. The reason is because FedCI accesses to all the data sources in a federated fashion while the `baselines trained locally' ($\mathsf{loc}$) only have access to a local data source.

\begin{figure}[!ht]
\centering
    \includegraphics[width=0.75\textwidth]{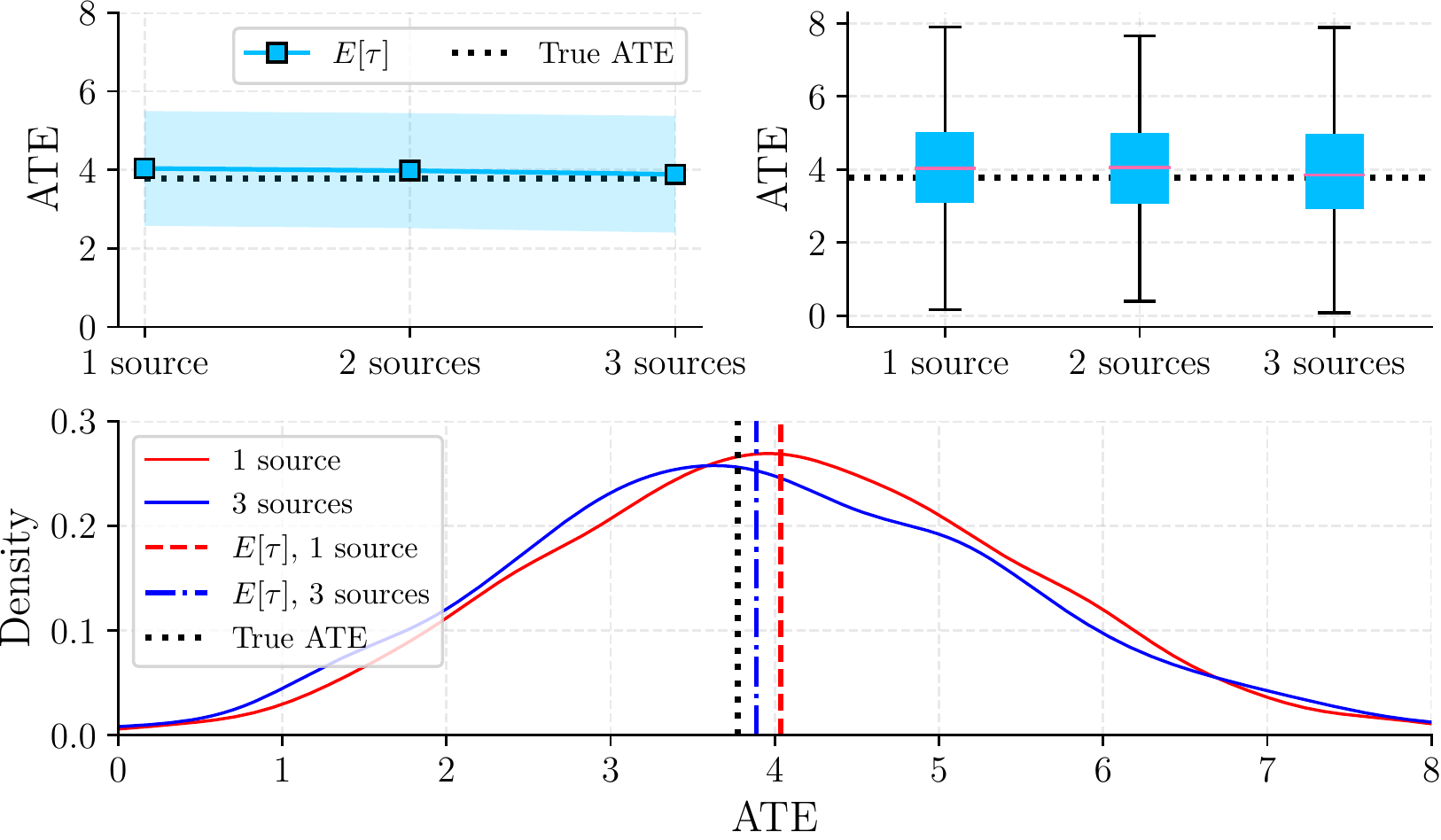}
    \vspace{-3pt}
\caption{The estimated ATE distribution on source \#1 of IHDP dataset. The dotted black lines represent the true ATE.}
\label{fig:fedci-uncertainty-analysis-ihdp}
\vspace{-6pt}
\end{figure}

Similar to the experiment on synthetic data, the estimated distribution of ATE in the first source ($\mathsf{s}=1$) is presented in Figure~\ref{fig:fedci-uncertainty-analysis-ihdp}. Again, the figures show that the true ATE is inside the estimated interval and the estimated mean ATE shifts towards its true value (dotted lines) when more data sources are used.

\begin{table}[!ht]
\centering
\caption{Out-of-sample errors on IHDP dataset where top-3 performances are highlighted in bold (lower is better). The dashes (---) in `$\mathsf{loc}$' and `$\mathsf{agg}$' indicate that the numbers are the same as those of `$\mathsf{com}$'.}
\vskip 3pt
\label{tab:error-ihdp-appendix}
\setlength{\tabcolsep}{2.7pt}
\small
\begin{tabular}{@{}lcccccc@{}}
\toprule
\multirow{2}{*}{Method}                                           & \multicolumn{3}{c}{The error of ITE ($\sqrt{\epsilon_\text{PEHE}}$)} & \multicolumn{3}{c}{The error of ATE ( $\epsilon_\text{ATE}$)} \\ \cmidrule(lr){2-4}\cmidrule(lr){5-7} 
                                                                  & 1 source      & 2 sources     & 3 sources     & 1 source      & 2 sources     & 3 sources    \\ \cmidrule(r){1-1}\cmidrule(lr){2-4}\cmidrule(lr){5-7}
BART$_\mathsf{loc}$                                                       & ---    & 5.83$\pm$2.6    & 6.56$\pm$3.3    & ---    & 2.09$\pm$0.9    & 1.38$\pm$0.5   \\
X-Learner$_\mathsf{loc}$                                                  & ---    & 4.14$\pm$1.5    & 4.54$\pm$1.9    & ---    & 1.51$\pm$0.7    & 0.77$\pm$0.5   \\
R-Learner$_\mathsf{loc}$                                                  & ---    & 6.35$\pm$1.9    & 6.16$\pm$2.0    & ---    & 2.13$\pm$0.7    & 1.44$\pm$0.3   \\
OthoRF$_\mathsf{loc}$                                                     & ---    & 4.33$\pm$1.6    & 4.59$\pm$1.9    & ---    & 1.10$\pm$0.6    & \textbf{0.75$\pm$0.3}   \\
CEVAE$_\mathsf{loc}$                                                     & ---    & 3.78$\pm$0.7    & 3.93$\pm$0.8    & ---    & 1.91$\pm$0.3    & 2.37$\pm$0.2   \\\cmidrule(r){1-1}\cmidrule(lr){2-4}\cmidrule(lr){5-7}
BART$_\mathsf{agg}$                                                 & ---             & 4.05$\pm$1.9              & \textbf{3.69$\pm$1.8}              & ---             & 2.09$\pm$1.0              & 1.30$\pm$0.5             \\
X-Learner$_\mathsf{agg}$ & ---             & 3.98$\pm$1.5    & 4.28$\pm$1.9    & ---             & 1.51$\pm$0.7    & 0.83$\pm$0.5   \\
R-Learner$_\mathsf{agg}$ & ---             & 4.76$\pm$1.3    & 4.46$\pm$1.6    & ---             & 1.92$\pm$0.5    & 1.41$\pm$0.2   \\
OthoRF$_\mathsf{agg}$    & ---             & \textbf{3.40$\pm$1.1}    & 4.26$\pm$1.9              & ---             & \textbf{0.87$\pm$0.3}    & 1.20$\pm$0.6             \\
CEVAE$_\mathsf{agg}$    & ---             & 3.63$\pm$0.7    & 3.73$\pm$0.5              & ---             & \textbf{0.92$\pm$0.2}    & 0.84$\pm$0.5             \\\cmidrule(r){1-1}\cmidrule(lr){2-4}\cmidrule(l){5-7}
BART$_\mathsf{com}$                                                 & 5.98$\pm$2.7             & 4.32$\pm$2.1              & 4.04$\pm$2.0              & \textbf{1.80$\pm$1.1}             & 2.09$\pm$1.1              & 1.21$\pm$0.6             \\
X-Learner$_\mathsf{com}$ & \textbf{4.22$\pm$1.6}             & 4.15$\pm$1.5    & 4.06$\pm$1.8    & \textbf{1.64$\pm$0.7}             & 1.93$\pm$0.8    & 0.84$\pm$0.4   \\
R-Learner$_\mathsf{com}$ & 6.97$\pm$2.1             & 4.43$\pm$1.4    & 4.47$\pm$1.7    & 3.15$\pm$0.5             & 1.34$\pm$0.5    & 1.10$\pm$0.3   \\
OthoRF$_\mathsf{com}$    & 4.49$\pm$1.9             & 3.81$\pm$1.3    & 3.75$\pm$1.5              & 1.86$\pm$0.8             & 1.61$\pm$0.6    & 1.56$\pm$0.8\\ 
CEVAE$_\mathsf{com}$        & \textbf{3.16$\pm$0.6}             & \textbf{2.34$\pm$0.6}    & \textbf{2.31$\pm$0.7}    & 2.02$\pm$0.4             & \textbf{0.53$\pm$0.1}    & \textbf{0.48$\pm$0.2}   \\\cmidrule(r){1-1}\cmidrule(lr){2-4}\cmidrule(lr){5-7}
FedCI                                                             & \textbf{2.88$\pm$0.8}     & \textbf{2.36$\pm$0.5}    & \textbf{2.35$\pm$0.6}    & \textbf{1.43$\pm$0.7}    & 1.03$\pm$0.4    & \textbf{0.51$\pm$0.2}   \\ \bottomrule
\end{tabular}
\end{table}

\subsection{Synthetic data: DATA-2}
In this section, we present additional  experimental results on DATA-2. Those results were skipped in the main text due to limited space. In Table~\ref{tab:error-synthetic-2-appendix} (five first rows), we present additional results of the baselines trained locally ($\mathsf{loc}$) and the baselines trained with bootstrap aggregating ($\mathsf{agg}$). Similar to the experiments on DATA-1 presented in the main text, the results on DATA-2 also show that FedCI achieves much lower errors, especially the error in predicting ITE.
\begin{table}[!ht]
\centering
\caption{Out-of-sample errors on synthetic dataset where top-3 performances are highlighted in bold (lower is better). The dashes (---) in `$\mathsf{loc}$' and `$\mathsf{agg}$' indicate that the numbers are the same as those of `$\mathsf{com}$'.}
\vskip 3pt
\label{tab:error-synthetic-2-appendix}
\setlength{\tabcolsep}{2.7pt}
\small
\begin{tabular}{@{}lcccccc@{}}
\toprule
\multirow{2}{*}{Method}                                           & \multicolumn{3}{c}{The error of ITE ($\sqrt{\epsilon_\text{PEHE}}$)} & \multicolumn{3}{c}{The error of ATE ( $\epsilon_\text{ATE}$)} \\ \cmidrule(lr){2-4}\cmidrule(lr){5-7} 
                                                                  & 1 source      & 3 sources     & 5 sources     & 1 source      & 3 sources     & 5 sources    \\ \cmidrule(r){1-1}\cmidrule(lr){2-4}\cmidrule(lr){5-7}
BART$_\mathsf{loc}$                                                       & ---    & 18.4$\pm$0.3    & 18.3$\pm$0.2    & ---    & 3.37$\pm$0.7    & 2.90$\pm$0.6   \\
X-Learner$_\mathsf{loc}$                                                  & ---    & 22.7$\pm$0.5    & 22.8$\pm$0.5    & ---    & 3.55$\pm$1.3    & 3.09$\pm$0.8   \\
R-Learner$_\mathsf{loc}$                                                  & ---    & 26.3$\pm$0.2    & 26.1$\pm$0.2    & ---    & 19.7$\pm$0.3    & 19.5$\pm$0.3   \\
OthoRF$_\mathsf{loc}$                                                     & ---   & 38.3$\pm$1.4    & 40.0$\pm$0.9    & ---    & 4.09$\pm$0.9    & 4.40$\pm$1.2   \\
CEVAE$_\mathsf{loc}$                                                     & ---   & $21.4\pm$0.7    & 19.8$\pm$0.6    & ---    & 2.11$\pm$0.4    & 1.97$\pm$0.2   \\\cmidrule(r){1-1}\cmidrule(lr){2-4}\cmidrule(lr){5-7}
BART$_\mathsf{agg}$                                                 & ---             & 17.9$\pm$0.2              & 17.7$\pm$0.2              & ---             & 3.91$\pm$0.8              & 3.15$\pm$0.7             \\
X-Learner$_\mathsf{agg}$ & ---             & 18.2$\pm$0.4    & 17.1$\pm$0.2    & ---             & 3.43$\pm$1.3    & 3.07$\pm$0.8   \\
R-Learner$_\mathsf{agg}$ & ---             & 26.2$\pm$0.3    & 26.1$\pm$0.2    & ---             & 19.7$\pm$0.4    & 19.6$\pm$0.3   \\
OthoRF$_\mathsf{agg}$    & ---             & 25.0$\pm$1.3    & 17.3$\pm$0.6              & ---             & 4.56$\pm$1.1    & \textbf{1.30$\pm$0.4}             \\
CEVAE$_\mathsf{agg}$                                                     & ---    & $19.2\pm$0.8    & 18.3$\pm$0.7    & ---    & 2.02$\pm$0.3    & 1.91$\pm$0.4   \\\cmidrule(r){1-1}\cmidrule(lr){2-4}\cmidrule(l){5-7}
BART$_\mathsf{com}$                                                     & \textbf{18.0$\pm$0.4}             & \textbf{17.7$\pm$0.2}    & 17.4$\pm$0.1    & \textbf{3.54$\pm$1.3}             & 2.94$\pm$0.8    & \textbf{1.84$\pm$0.5}   \\
X-Learner$_\mathsf{com}$    & 21.1$\pm$0.9             & 17.9$\pm$0.4    & \textbf{16.2$\pm$0.2}    & 4.55$\pm$1.4             & 3.29$\pm$1.0    & 2.37$\pm$0.8   \\
R-Learner$_\mathsf{com}$     & 25.9$\pm$0.6             & 23.5$\pm$0.5    & 21.3$\pm$0.4    & 19.0$\pm$0.8             & 15.6$\pm$0.7    & 12.3$\pm$0.6   \\
OthoRF$_\mathsf{com}$        & 37.8$\pm$2.7             & \textbf{10.7$\pm$0.5}    & \textbf{9.83$\pm$0.5}    & 7.88$\pm$2.2             & \textbf{1.99$\pm$0.4}    & 2.36$\pm$0.6   \\
CEVAE$_\mathsf{com}$        & \textbf{20.1$\pm$0.5}             & 18.4$\pm$0.6    & 16.6$\pm$0.6    & \textbf{1.50$\pm$0.3}             & \textbf{1.38$\pm$0.4}    & 1.89$\pm$0.2   \\\cmidrule(r){1-1}\cmidrule(lr){2-4}\cmidrule(lr){5-7}
FedCI                                                             & \textbf{9.28$\pm$0.4}     & \textbf{6.34$\pm$0.2}    & \textbf{5.53$\pm$0.1}    & \textbf{2.37$\pm$0.5}    & \textbf{1.47$\pm$0.4}    & \textbf{0.74$\pm$.2}   \\ \bottomrule
\end{tabular}
\end{table}

\section{Proof of Lemma 1}
\begin{proof}
\begin{align*}
\mathbb{C}\text{ov}(\mathbf{y}_i^{\mathsf{s}}, \mathbf{y}_j^{\mathsf{s}'}\,|\,\Sigma,\Phi) &= \mathbb{C}\text{ov}\Big(\Phi^{\frac{1}{2}} (\mathbf{f}_i^{\mathsf{s}} + \mathbf{g}^{\mathsf{s}}) + \Sigma^{\frac{1}{2}}\bm{\varepsilon}_i^{\mathsf{s}} , \Phi^{\frac{1}{2}} (\mathbf{f}_j^{\mathsf{s}'} + \mathbf{g}^{\mathsf{s}'}) + \Sigma^{\frac{1}{2}}\bm{\varepsilon}_j^{\mathsf{s}'} \,|\,\Sigma,\Phi\Big)\\
&= \e\Big[\Phi^{\frac{1}{2}}\mathbf{g}^{\mathsf{s}}  (\mathbf{g}^{\mathsf{s}'})^\top(\Phi^{\frac{1}{2}})^\top \Big] - \Phi^{\frac{1}{2}}\e\Big[\mathbf{g}^{\mathsf{s}} \Big]\e\Big[ \mathbf{g}^{\mathsf{s}'} \Big]^\top\Phi^{\frac{1}{2}})^\top\\
&= \Phi^{\frac{1}{2}}\Lambda^{(\mathsf{s},\mathsf{s}')}(\Phi^{\frac{1}{2}})^\top
\end{align*}

This completes the proof. 
\end{proof}

\section{Proof of Lemma 2}

\begin{proof}
We denote $\xi_0^{\mathsf{s}} \sim \mathsf{N}(\mathbf{0}, \mathbf{I}_{n_{\mathsf{s}}})$ and $\xi_1^{\mathsf{s}} \sim \mathsf{N}(\mathbf{0}, \mathbf{I}_{n_{\mathsf{s}}})$. Then, from the model definition (Eq.~(5) in the main text), we have
\begin{align*}
&\setlength\arraycolsep{1.0pt} \begin{bmatrix}
y_1^{\mathsf{s}}(0)&\dots&y_{n_{\mathsf{s}}}^{\mathsf{s}}(0)\\
y_1^{\mathsf{s}}(1)&\dots&y_{n_{\mathsf{s}}}^{\mathsf{s}}(1)
\end{bmatrix}
\setlength\arraycolsep{1.0pt}\!=\!\Phi^{\frac{1}{2}} \begin{bmatrix}
f_1^{\mathsf{s}}(0)+g^{\mathsf{s}}(0)&\dots&f_{n_{\mathsf{s}}}^{\mathsf{s}}(0)+g^{\mathsf{s}}(0)\\
f_1^{\mathsf{s}}(1)+g^{\mathsf{s}}(1)&\dots&f_{n_{\mathsf{s}}}^{\mathsf{s}}(1)+g^{\mathsf{s}}(1)
\end{bmatrix} 
\setlength\arraycolsep{1.0pt}\!+\!\Sigma^{\frac{1}{2}}\begin{bmatrix}
\varepsilon_1^{\mathsf{s}}(0)&\dots&\varepsilon_{n_{\mathsf{s}}}^{\mathsf{s}}(0)\\
\varepsilon_1^{\mathsf{s}}(1)&\dots&\varepsilon_{n_{\mathsf{s}}}^{\mathsf{s}}(1)
\end{bmatrix},
\end{align*}

which is equivalent to the following
\begin{align*}
&\mathbf{Y}^{\mathsf{s}} \!=\!  \setlength\arraycolsep{5pt} \begin{bmatrix}
\mu_0(\mathbf{X}^{\mathsf{s}}) \!+\! \mathbf{g}_0^{\mathsf{s}} \!+\! (\mathbf{K}^{\mathsf{s}})^{\frac{1}{2}}\xi_0^{\mathsf{s}}&\mu_1(\mathbf{X}^{\mathsf{s}}) \!+\! \mathbf{g}_1^{\mathsf{s}} \!+\! (\mathbf{K}^{\mathsf{s}})^{\frac{1}{2}}\xi_1^{\mathsf{s}}
\end{bmatrix}(\Phi^{\frac{1}{2}})^\top \!+\! \setlength\arraycolsep{2pt} \begin{bmatrix}
\varepsilon_0^{\mathsf{s}}&\varepsilon_1^{\mathsf{s}}
\end{bmatrix}(\Sigma^{\frac{1}{2}})^\top\\
&\mathbf{Y}^{\mathsf{s}} \!=\! \setlength\arraycolsep{2pt}\begin{bmatrix}\mu_0(\mathbf{X}^{\mathsf{s}})\!+\!\mathbf{g}_0^{\mathsf{s}}&\mu_1(\mathbf{X}^{\mathsf{s}})+\mathbf{g}_1^{\mathsf{s}}
\end{bmatrix}(\Phi^{\frac{1}{2}})^\top \!+\! (\mathbf{K}^{\mathsf{s}})^{\frac{1}{2}}\setlength\arraycolsep{3pt}\begin{bmatrix}\xi_0^{\mathsf{s}}&\xi_1^{\mathsf{s}}
\end{bmatrix}(\Phi^{\frac{1}{2}})^\top\!+\! \setlength\arraycolsep{2pt}\begin{bmatrix}
\varepsilon_0^{\mathsf{s}}&\varepsilon_1^{\mathsf{s}}
\end{bmatrix}(\Sigma^{\frac{1}{2}})^\top\\
&\text{vec}(\mathbf{Y}^{\mathsf{s}}) =  \left(\Phi^{\frac{1}{2}} \otimes \mathbf{I}_{n_{\mathsf{s}}}\right)\begin{bmatrix}\mu_0(\mathbf{X}^{\mathsf{s}})+\mathbf{g}_0^{\mathsf{s}}\\\mu_1(\mathbf{X}^{\mathsf{s}})+\mathbf{g}_1^{\mathsf{s}}
\end{bmatrix}  +\left(\Phi^{\frac{1}{2}} \otimes (\mathbf{K}^{\mathsf{s}})^{\frac{1}{2}}\right)\begin{bmatrix}\xi_0^{\mathsf{s}}\\\xi_1^{\mathsf{s}}
\end{bmatrix} + (\Sigma^{\frac{1}{2}} \otimes \mathbf{I}_{n_{\mathsf{s}}})\begin{bmatrix}
\varepsilon_0^{\mathsf{s}}\\\varepsilon_1^{\mathsf{s}}
\end{bmatrix},
\end{align*}
where $\text{vec}(\cdot)$ denotes the vectorization of a matrix, which converts a matrix into a column vector.

For the second term on the right hand side of the above equation, note that $\xi_0^{\mathsf{s}} \sim \mathsf{N}(\mathbf{0}, \mathbf{I}_{n_{\mathsf{s}}})$ and $\xi_1^{\mathsf{s}} \sim \mathsf{N}(\mathbf{0}, \mathbf{I}_{n_{\mathsf{s}}})$, so we have the following
\begin{align*}
&\begin{bmatrix}\xi_0^{\mathsf{s}}\\\xi_1^{\mathsf{s}}
\end{bmatrix} \sim \mathsf{N}(\mathbf{0}, \mathbf{I}_{2n_{\mathsf{s}}})\\
&\left(\Phi^{\frac{1}{2}} \otimes (\mathbf{K}^{\mathsf{s}})^{\frac{1}{2}}\right)\begin{bmatrix}\xi_0^{\mathsf{s}}\\\xi_1^{\mathsf{s}}
\end{bmatrix} \sim \mathsf{N}\left(\mathbf{0}, \left(\Phi^{\frac{1}{2}} \otimes (\mathbf{K}^{\mathsf{s}})^{\frac{1}{2}}\right)\mathbf{I}_{2N}\left(\Phi^{\frac{1}{2}} \otimes (\mathbf{K}^{\mathsf{s}})^{\frac{1}{2}}\right)^\top\right)\\
&\left(\Phi^{\frac{1}{2}} \otimes (\mathbf{K}^{\mathsf{s}})^{\frac{1}{2}}\right)\begin{bmatrix}\xi_0^{\mathsf{s}}\\\xi_1^{\mathsf{s}}
\end{bmatrix} \sim \mathsf{N}\left(\mathbf{0}, \Phi\otimes \mathbf{K}^{\mathsf{s}}\right).
\end{align*}
\vspace{-6pt}
For the last term, note that $\varepsilon_0^{\mathsf{s}} \sim \mathsf{N}(0, \mathbf{I}_{n_{\mathsf{s}}}), \varepsilon_1^{\mathsf{s}} \sim \mathsf{N}(0, \mathbf{I}_{n_{\mathsf{s}}})$, thus
\begin{align*}
&\begin{bmatrix}\varepsilon_0^{\mathsf{s}}\\\varepsilon_1^{\mathsf{s}}
\end{bmatrix} \sim \mathsf{N}(\mathbf{0}, \mathbf{I}_{2n_{\mathsf{s}}})\\
&\left(\Sigma^{\frac{1}{2}} \otimes \mathbf{I}_{n_{\mathsf{s}}}\right)\begin{bmatrix}\varepsilon_0^{\mathsf{s}}\\\varepsilon_1^{\mathsf{s}}
\end{bmatrix} \sim \mathsf{N}\left(\mathbf{0}, \left(\Sigma^{\frac{1}{2}} \otimes \mathbf{I}_{n_{\mathsf{s}}}\right)\mathbf{I}_{2n}\left(\Sigma^{\frac{1}{2}} \otimes \mathbf{I}_{n_{\mathsf{s}}}\right)^\top\right)\\
&\left(\Sigma^{\frac{1}{2}} \otimes \mathbf{I}_{n_{\mathsf{s}}}\right)\begin{bmatrix}\varepsilon_0^{\mathsf{s}}\\\varepsilon_1^{\mathsf{s}}
\end{bmatrix} \sim \mathsf{N}\left(\mathbf{0}, \Sigma\otimes \mathbf{I}_{n_{\mathsf{s}}}\right).
\end{align*}
Consequently, 
\vspace{-6pt}
\begin{align*}
\text{vec}(\mathbf{Y}^{\mathsf{s}}) \big| \Phi, \Sigma, \mathbf{X}^{\mathsf{s}}, \mathbf{w}^{\mathsf{s}}, \mathbf{g}^{\mathsf{s}} \sim \mathsf{N}\left( \left(\Phi^{\frac{1}{2}} \otimes \mathbf{I}_{n_{\mathsf{s}}}\right)\begin{bmatrix}\mu_0(\mathbf{X}^{\mathsf{s}})\!+\! \mathbf{g}_0^{\mathsf{s}}\\\mu_1(\mathbf{X}^{\mathsf{s}}) \!+\! \mathbf{g}_1^{\mathsf{s}}
\end{bmatrix} \!, \Phi \otimes \mathbf{K}^{\mathsf{s}} \!+\! \Sigma \otimes \mathbf{I}_{n_{\mathsf{s}}}\right)\!,
\end{align*}
which implies that
\begin{align*}
\begin{bmatrix}
\mathbf{y}^{\mathsf{s}}(0)\\
\mathbf{y}^{\mathsf{s}}(1)
\end{bmatrix}\Big|\Phi, \Sigma, \mathbf{X}^{\mathsf{s}}, \mathbf{w}^{\mathsf{s}}, \mathbf{g}^{\mathsf{s}} \sim \mathsf{N}\left( \left(\Phi^{\frac{1}{2}} \otimes \mathbf{I}_{n_{\mathsf{s}}}\right)\begin{bmatrix}\mu_0(\mathbf{X}^{\mathsf{s}})\!+\! \mathbf{g}_0^{\mathsf{s}}\\\mu_1(\mathbf{X}^{\mathsf{s}})\!+\! \mathbf{g}_1^{\mathsf{s}}
\end{bmatrix} \!, \Phi \otimes \mathbf{K}^{\mathsf{s}} \!+\! \Sigma \otimes \mathbf{I}_{n_{\mathsf{s}}}\right)\!.
\end{align*}
This completes the proof.
\end{proof}

\section{Proof of Lemma 3}

\begin{proof}
Following the proof of Lemma~\ref{lem:joint-prob-s}, we note that if the observed treatment $w_i^{\mathsf{s}} = 0$, then the mean of $p(y^{\mathsf{s}}_{i,\textrm{obs}} | \mathbf{X}^{\mathsf{s}}, \mathbf{w}^{\mathsf{s}}, \Phi, \Sigma, \mathbf{g}^{\mathsf{s}})$ equals to the mean of $p(y_i^{\mathsf{s}}(0) | \Phi, \Sigma, \mathbf{X}^{\mathsf{s}}, \mathbf{w}^{\mathsf{s}}, \mathbf{g}^{\mathsf{s}})$ and the mean of $p(y^{\mathsf{s}}_{i,\textrm{mis}} | \mathbf{X}^{\mathsf{s}}, \mathbf{w}^{\mathsf{s}}, \Phi, \Sigma, \mathbf{g}^{\mathsf{s}})$ equals to the mean of $p(y_i^{\mathsf{s}}(1) | \Phi, \Sigma, \mathbf{X}^{\mathsf{s}}, \mathbf{w}^{\mathsf{s}}, \mathbf{g}^{\mathsf{s}})$. If the observed treatment $w_i^{\mathsf{s}} = 1$, then the mean of $p(y^{\mathsf{s}}_{i,\textrm{obs}} | \mathbf{X}^{\mathsf{s}}, \mathbf{w}^{\mathsf{s}}, \Phi, \Sigma, \mathbf{g}^{\mathsf{s}})$ equals to the mean of $p(y_i^{\mathsf{s}}(1) | \Phi, \Sigma, \mathbf{X}^{\mathsf{s}}, \mathbf{w}^{\mathsf{s}}, \mathbf{g}^{\mathsf{s}})$ and the mean of $p(y^{\mathsf{s}}_{i,\textrm{mis}} | \mathbf{X}^{\mathsf{s}}, \mathbf{w}^{\mathsf{s}}, \Phi, \Sigma, \mathbf{g}^{\mathsf{s}})$ equals to the mean of $p(y_i^{\mathsf{s}}(0) | \Phi, \Sigma, \mathbf{X}^{\mathsf{s}}, \mathbf{w}^{\mathsf{s}}, \mathbf{g}^{\mathsf{s}})$. Similarly, each element in $\mathbf{K}_{\textrm{obs}}$ and  $\mathbf{K}_{\textrm{mis}}$ also depends on whether $w_i^{\mathsf{s}} = 0$ or $w_i^{\mathsf{s}} = 1$.
\end{proof}

\end{document}